\documentclass[12pt]{IEEEtran} 
\usepackage{graphicx}
\usepackage{bm,amsmath}
\usepackage{float}

\def\BibTeX{{\rm B\kern-.05em{\sc i\kern-.025em b}\kern-.08em
    T\kern-.1667em\lower.7ex\hbox{E}\kern-.125emX}}

\newtheorem{theorem}{Theorem}
\begin{document}

\title{On the Capacity of Multiple Antenna Systems in Rician Fading} 

\author{Sudharman K. Jayaweera and H. Vincent Poor \thanks{Sudharman K.
Jayaweera was with the Department of Electrical Engineering,
Princeton University, Princeton, NJ, 08544, USA. He is now with
the Department of Electrical and Computer Engineering, Wichita
State University, Wichita, KS, 67226, USA. E-mail:
sjayawee@ee.princeton.edu.}
\thanks{H. Vincent Poor is with the Department of Electrical
Engineering, Princeton University, Princeton, NJ, 08544, USA.
E-mail: poor@princeton.edu.}
\thanks{This research was supported in part by the National
Science Foundation under Grant $CCR-0086017$, and in part by the
New Jersey Center for Wireless Telecommunications.}}


\thispagestyle{empty} \pagestyle{plain} \maketitle

\begin{abstract}
The effect of {\it Rician-ness} on the capacity of multiple
antenna systems is investigated under the assumption that channel
state information (CSI) is available only at the receiver. The
average-power-constrained capacity of such systems is considered
under two different assumptions on the knowledge about the fading
available at the transmitter: the case in which the transmitter
has no knowledge of fading at all, and the case in which the
transmitter has knowledge of the distribution of the fading
process but not the instantaneous CSI. The exact capacity is given
for the former case while capacity bounds are derived for the
latter case. A new signalling scheme is also proposed for the
latter case and it is shown that by exploiting the knowledge of
{\it Rician-ness} at the transmitter via this signalling scheme,
significant capacity gain can be achieved. The derived capacity
bounds are evaluated explicitly to provide numerical results in
some representative situations.
\end{abstract}


\section{Introduction}

Prompted by recent results suggesting possible extraordinary
capacity gains \cite{Foschini1,FoschiniGans1,Telatar1}, multiple
transmit and receive antenna systems have received considerable
attention as a means of providing substantial performance
improvement in wireless communication systems. In such
multiple-input/multiple-output (MIMO) systems, multiple
transmit/receiver antenna combinations  provide spatial diversity
by exploiting channel fading. It has been shown in
\cite{FoschiniGans1,Telatar1} that when the receiver has access to
perfect channel state information but not the transmitter, the
capacity of a Rayleigh distributed flat fading channel will
increase almost linearly with the minimum of the number of
transmit and receive antennas.

In most previous research on the capacity of multiple antenna
systems, however, the channel fading is assumed to be Rayleigh
distributed. Of course, the Rayleigh fading model is known to be a
reasonable assumption for fading encountered in many wireless
communications systems. However, it is also of interest to
investigate the capacity of multiple antenna systems when the
Rayleigh fading model is replaced by the more general Rician
model\index{Rician distribution}. Not only does this generalize
the previously derived capacity results, since both additive white
Gaussian noise (AWGN) and Rayleigh fading channels may be
considered to be limiting cases of the Rician channel, but Rician
fading is also known to be a better model for wireless
environments with a strong direct Line-Of-Sight (LOS) path
\cite{Stuber1}. In this paper, we consider the capacity of
multiple antenna systems in Rician fading for two cases of
interest: that in which the receiver has perfect channel state
information (CSI) but the transmitter has no knowledge of the
fading statistics; and that in which the receiver has perfect CSI
and the transmitter knows the distribution of the fading process,
but not exact CSI.

We begin, in Section \ref{sec:rc_model}, by introducing the
multiple antenna system model of interest and the assumptions on
the fading process. Next, in Section \ref{sec:rc_RiceCap1} we
address the general capacity problem for the Rician fading channel
and obtain an upper bound for the capacity of multiple antenna
systems under Rician fading. We explicitly evaluate this upper
bound for some special cases.

In Section \ref{sec:rc_RiceCap2} we investigate the exact capacity
of MIMO systems in Rician channels under the assumptions of
perfect channel state information at the receiver and no knowledge
of the fading distribution at the transmitter. In addition to
providing a lower bound on the capacity for the case in which the
transmitter does know the fading distribution, this result also
serves as a measure of the capacity variation of a system designed
under Rayleigh fading assumption but operating in an environment
where a strong LOS component is present. This is because the
capacity-achieving distribution for the case considered in this
section also achieves the capacity in the Rayleigh channel. We
explicitly evaluate this capacity for some interesting special
cases.

We will see that there is a large capacity gap between the upper
bound of Section \ref{sec:rc_RiceCap1} and the lower bound of
Section \ref{sec:rc_RiceCap2} obtained with signals designed to be
optimal for Rayleigh fading. In Section \ref{sec:rc_newSigs} we
propose a new signalling scheme for multiple antenna systems with
perfect-CSI at the receiver and only knowledge of the fading
distribution, but not the exact CSI, at the transmitter. We derive
tight upper and lower bounds for the capacity of a multiple
transmit antenna system with this new input signal choice. By
comparing these capacity bounds with the results obtained in
Section \ref{sec:rc_RiceCap2}, we will show that the exploitation
of the knowledge of the fading distribution at the transmitter can
provide significant capacity gains. Finally, we finish with some
concluding remarks in Section \ref{sec:rc_conclusions}.

Some mathematical results that we will need in the rest of the
paper are given in the Appendix.

\section{Model Description} \label{sec:rc_model}

We consider a single user, narrowband, MIMO communication link in
which the transmitter and receiver are equipped with $N_T$ and
$N_R$ antennas, respectively. We consider the ideal case in which
the antenna elements at both transmitter and receiver are
sufficiently far apart so that the fading corresponding to
different antenna elements is uncorrelated. The discrete-time
received signal in such a system can be written in matrix form as
\begin{eqnarray}
{\bf y}(i) &=& {\bf H}(i) {\bf x}(i) + {\bf n}(i) \ ,
\label{eq:rc_y_i}
\end{eqnarray}
where ${\bf y}(i)$, ${\bf x}(i)$ and ${\bf n}(i)$ are the complex
$N_R$-vector of received signals on the $N_R$ receive antennas,
the (possibly) complex $N_T$-vector of transmitted signals on the
$N_T$ transmit antennas, and the complex $N_R$-vector of additive
receiver noise, respectively, at symbol time $i$. The components
of ${\bf n}(i)$ are independent, zero-mean, circularly symmetric
complex Gaussian random variables with independent real and
imaginary parts having equal variances; i.e. ${\bf n}(i) \sim
{\cal N}_c \left( {\bf 0}, {\bf I}_{N_R} \right)$, where ${\bf
I}_{N_R}$ denotes the $N_R \times N_R$ identity matrix. The noise
is also assumed to be independent with respect to the time index.

The matrix ${\bf H}(i)$ in the model (\ref{eq:rc_y_i}) is the $N_R
\times N_T$ matrix of complex fading coefficients. The
$(n_R,n_T)$-th element of the matrix ${\bf H}(i)$, denoted by
$\left({\bf H}(i) \right)_{n_R,n_T}$, represents the fading
coefficient value at time $i$ between the $n_R$-th receiver
antenna and the $n_T$-th transmitter antenna. The fading
coefficients in each channel use are considered to be independent
from those of other channel uses, i.e. $\left\{ {\bf H}(i)
\right\}$ is an independent sequence. As noted in \cite{Telatar1},
this gives rise to a memoryless channel, and thus the capacity of
the channel can be computed as the maximum mutual
information\index{mutual information},
\begin{eqnarray}
C_{N_T,N_R} &=&  \max_{P_{\bf X}} \ \ {\cal I} ({\bf x}; {\bf y})
\nonumber
\end{eqnarray}
where $P_{\bf X}$ is the probability distribution of the input
signal vector ${\bf x}$ that satisfies a given power constraint at
the transmitter and ${\cal I} ({\bf x}; {\bf y})$ is the mutual
information between the input ${\bf x}$ and output ${\bf y}$. In
this case, we may also drop the explicit time index, $i$, in order
to simplify notation.

The main purpose of this paper is to extend the previously known
capacity results for multiple antenna systems in Rayleigh fading
to Rician channels. Thus, we will assume that the elements of
${\bf H}$ are Gaussian with independent real and imaginary parts
each distributed as ${\cal N}(\mu/ \sqrt{2}, \sigma^2)$. Moreover,
the elements of ${\bf H}$ are assumed to be independent of each
other. So, the elements $\left({\bf H} \right)_{n_R,n_T}$ of ${\bf
H}$ are independent and identically distributed (i.i.d.) complex
Gaussian random variables $\left({\bf H} \right)_{n_R,n_T} \sim
{\cal N}_c \left( \frac{\mu}{\sqrt{2}}(1+j), 2 \sigma^2 \right)$,
for $n_R = 1, \ldots, N_R$ and $n_T = 1, \ldots , N_T$, and the
distribution of the magnitudes of the elements of ${\bf H}$ have
the following Rician probability density function (pdf):
\begin{eqnarray}
f_R(r) &=& 2(1+\kappa) r e^{-(1+\kappa) r^2 - \kappa}
I_0(2\sqrt{\kappa (1+\kappa)}r) , \label{eq:rc_rician2}
\end{eqnarray}
where $I_0$ is the zero'th order modified Bessel function of the
first kind \cite{McLachlan1} and we have introduced the Rician
factor, $\kappa$, defined as
\begin{equation}
\kappa = \frac{|\mu|^2}{2 \sigma^2} . \label{eq:rc_rice factor1}
\end{equation}
For notational convenience, we have also introduced the
normalization $|\mu|^2 + 2 \sigma^2 = 1$. Note that
(\ref{eq:rc_rician2}) reduces to the Rayleigh pdf when $\kappa =
0$ (which implies that $\mu=0$).

When elements of ${\bf H}$ are distributed as described above we
say that ${\bf H}$ is a complex normally distributed matrix,
denoted as ${\bf H} \sim {\cal N}_c \left( {\bf M}, {\bf I}_{N_T}
\otimes {\bm \Sigma} \right)$ where ${\bm \Sigma}$ is the
Hermitian covariance matrix of the columns (assumed to be the same
for all columns) of ${\bf H}$ and ${\bf M} = \emph{E} \left\{ {\bf
H} \right\} $. For the assumed model,
\begin{eqnarray}
{\bm \Sigma} &=& 2 \sigma^2 {\bf I}_{N_R} , \label{eq:rc_Cov1}
\end{eqnarray}
and
\begin{eqnarray}
{\bf M} &=&  \frac{\mu}{\sqrt{2}} (1+ j) {\bm \Psi}_{N_R,N_T} ,
\label{eq:rc_Mean1}
\end{eqnarray}
where ${\bm \Psi}_{N_R, N_T}$ denotes the $N_R \times N_T$ matrix
of all ones.

Next, let us define $n = \max \{ N_R, N_T \}$, $m = \min \{ N_R,
N_T \}$ and
\begin{eqnarray}
{\bf W} &=& \left\{ \begin{array}{cc}
{\bf H} {\bf H}^H & \textrm{if $N_R < N_T$} \\
{\bf H}^H{\bf H} & \textrm{if $N_R \geq N_T$} \end{array} \right.
. \label{eq:rc_W}
\end{eqnarray}

Then, ${\bf W}$ is always an $m \times m$ square matrix. It is
known that when ${\bf H}$ is a complex normally distributed matrix
as described above, the distribution of ${\bf W}$ is given by the
non-central Wishart distribution\index{Wishart
distribution!non-central} \cite{Constantine1,James1,James2} with
pdf
\begin{eqnarray}
f_{{\bf W}}({\bf W}) &=& e^{-{\textrm{tr}} \left\{ {\bm
\Sigma}^{-1}{\bf M} {\bf M}^H \right\}} {}_0\tilde{F}_1 (n;{\bm
\Sigma}^{-1}{\bf M} {\bf M}^H {\bm \Sigma}^{-1} {\bf W}) f_{{\bf
W}}^0({\bf W}) , \label{eq:rc_nonCentrlWishart2}
\end{eqnarray}
where in (\ref{eq:rc_nonCentrlWishart2}) $f_{{\bf W}}^0({\bf W})$
denotes the (central) Wishart pdf\index{Wishart distribution}:
\begin{eqnarray}
f_{{\bf W}}^0({\bf W}) &=& \frac{1}{\tilde{\Gamma}_{m}(n) |{\bm
\Sigma}|^n} e^{-{\textrm{tr}} {\bm \Sigma}^{-1} {\bf W}} |{\bf
W}|^{n-m}  , \label{eq:rc_CentrlWishart1}
\end{eqnarray}
which results when the elements of ${\bf H}$ are iid zero mean
Gaussian random variables, and where the complex multivariate
gamma function $\tilde{\Gamma}_{m}$ and the Bessel
function\index{Bessel function!matrix} of matrix argument
${}_0\tilde{F}_1$ are defined in the Appendix (see
(\ref{eq:rc_multiGammaComplex2}) and
(\ref{eq:rc_BesselFuncOfMatrix1}))\footnote{Note that
(\ref{eq:rc_BesselFuncOfMatrix1}) applies in this case by noting
that ${\bm \Sigma}^{-1}{\bf M} {\bf M}^H {\bm \Sigma}^{-1} {\bf W}
= {\bm \Sigma}^{-1}{\bf M} {\bf M}^H {\bm \Sigma}^{-1} {\bf H}
{\bf H}^H$ and the trace relationship ${\textrm{tr}} \left( {\bf
A} {\bf B} \right) = {\textrm{tr}} \left( {\bf B} {\bf
A}\right)$.}.

Note that in (\ref{eq:rc_nonCentrlWishart2}) we have assumed,
without loss of generality, that ${\bf W}={\bf H} {\bf H}^H$. We
will continue to use this assumption throughout unless stated
otherwise. We use the shorthand notations ${\bf W} \sim {\cal W}_m
\left(n, {\bm \Sigma} \right)$ and ${\bf W} \sim {\cal W}_m
\left(n, {\bm \Sigma}, {\bm \Sigma}^{-1}{\bf M} {\bf M}^H \right)$
to denote that ${\bf W}$ has the Wishart distribution with pdf
(\ref{eq:rc_CentrlWishart1}) and that ${\bf W}$  has the
non-central Wishart distribution with pdf
(\ref{eq:rc_nonCentrlWishart2}), respectively.

\section{Capacity of the Multiple Antenna Rician Fading Channel}
\label{sec:rc_RiceCap1}

It is shown in \cite{Telatar1} that, for the Rayleigh flat fading
channel (i.e. the model of Section \ref{sec:rc_model} with
$\kappa=0$) under the total average power constraint $\emph{E} \{
{\bf x}^H {\bf x} \} \leq P$, the capacity of the channel
(\ref{eq:rc_y_i}) is achieved when ${\bf x}$ has a circularly
symmetric complex Gaussian distribution with zero-mean and
covariance $ \frac{P}{N_T} {\bf I}_{N_T}$, and that this capacity
is given by the expression $C_{N_T,N_R}^0 = \emph{E} \left\{
\log\det \left( {\bf I}_{N_R} + \frac{P}{N_T} {\bf H}{\bf H}^H
\right) \right\}$.

However, it is also easily shown that the capacity achieving
transmit signal distribution ${\bf x}$ for a multiple antenna
system under the average power constraint above is circularly
symmetric, zero-mean complex Gaussian regardless of the actual
fading distribution, as long as the receiver, but not the
transmitter, knows the channel fading coefficients. Thus, only the
covariance matrix ${\bf Q} = \emph{E} \{ {\bf x} {\bf x}^H \}$ of
the capacity-achieving distribution depends on the fading
distribution, and the capacity of the multiple antenna system is
then given by
\begin{eqnarray}
C_{N_T,N_R} &=&   {\underset{ {\underset{ {\bf Q} \ \geq \
0}{{\textrm{tr}} \ {\bf Q} \ \leq \ P}} }{\max}} \ \
\emph{E}_{{\bf H}} \{ \log \det ( {\bf H} {\bf Q} {\bf H}^H + {\bf
I}_{N_R} ) \} \ . \label{eq:rc_capacity_RicianGeneral}
\end{eqnarray}
In the case of deterministic fading where the matrix ${\bf H}$ has
all its elements equal to unity (i.e. the Rician model with
$\kappa \longrightarrow \infty$) and this is known to the
transmitter, the so called water-filling\index{water-filling}
algorithm \cite{gallager1} specifies the covariance matrix
structure of this capacity achieving Gaussian distribution to be
of the form ${\bf Q}^{\infty} =  \frac{P}{N_T} {\bm \Psi}_{N_T}$
where ${\bm \Psi}_{N_T}$ denotes the $N_T \times N_T$ matrix of
all ones \cite{Telatar1}. In this case, the capacity is given by
\begin{eqnarray}
C_{N_T,N_R}^{\infty} &=& \log(1 + N_R N_T P) .
\label{eq:rc_capacity_RicianGeneral11}
\end{eqnarray}
Alternatively, in the case where the fading is Rician but without
knowledge of $\kappa$ at the transmitter, the capacity achieving
distribution is the same as in the Rayleigh case
\cite{Foschini1,FoschiniGans1,Telatar1}, i.e. its covariance
matrix is ${\bf Q}^{0} =  \frac{P}{N_T} {\bf I}_{N_T}$.

Thus, for a channel with Rician distributed fading having a
general value of $\kappa$, which is known to the transmitter, one
would expect the covariance matrix ${\bf Q}$ of the capacity
achieving distribution to lie in between these two extremes.
Although the capacity-achieving ${\bf Q}$ for this case is
unknown, in the following paragraphs we derive an upper bound for
the capacity of this channel.

Observe that for any ${\bf Q}$ the matrix ${\bf H} {\bf Q} {\bf
H}^H + {\bf I}_{N_R}$ is positive definite, and that the function
$\log \det$ is concave on the set of positive definite matrices.
Thus, applying Jensen's inequality\index{Jensen's inequality} to
(\ref{eq:rc_capacity_RicianGeneral}) we have
\begin{eqnarray}
C_{N_T,N_R} &\leq&  {\underset{ {\underset{ {\bf Q} \ \geq \
0}{{\textrm{tr}} \ {\bf Q} \ \leq \ P}} }{\max}} \ \ \log \det
\left(N_R {\bf Q} {\bm \Upsilon} + {\bf I}_{N_T} \right) ,
\label{eq:rc_capacity_RicianGeneral4}
\end{eqnarray}
where we have used the determinant identity $\det(I + A B) = \det
(I + B A)$ and introduced the notation
\begin{eqnarray}
\emph{E} \{ {\bf H}^H {\bf H} \} &=& N_R {\bm \Upsilon} .
\end{eqnarray}

It is easy to show that the $N_T \times N_T$ matrix ${\bm
\Upsilon}$ is given by
\begin{eqnarray}
{\bm \Upsilon} &=& \frac{1}{1+ \kappa} \left[ \begin{array}{cccc}
1+\kappa & \kappa   & \ldots & \kappa \\
\kappa   & 1+\kappa & \ldots & \kappa \\
\vdots   & \vdots   & \ddots & \vdots \\
\kappa   & \kappa   & \ldots & 1+\kappa
\end{array} \right] . \label{eq:rc_ExpectMat1}
\end{eqnarray}

We observe that for any $\kappa$ such that $0 \leq \kappa <
\infty$ the matrix ${\bm \Upsilon}$ is non-singular and thus all
the eigenvalues of ${\bm \Upsilon}$ are non-zero. In fact, if we
denote the eigenvalues of ${\bm \Upsilon}$ by $\lambda_i$ for $i =
1, \ldots , N_T$, then it can be shown that,
\begin{eqnarray}
{\lambda_i} &=& \left\{ \begin{array}{cc}
\frac{1 + N_T \kappa}{ 1+ \kappa} & \textrm{if $i = 1$} \\
\frac{1}{ 1+ \kappa} & \textrm{if $i = 2, \ldots , N_T$}
\end{array} \right. \ \ \ {\textrm{for}} \ \ \ 0 \leq \kappa <
\infty. \label{eq:rc_eigs1}
\end{eqnarray}

We may decompose ${\bm \Upsilon}$ as ${\bm \Upsilon} = {\bf U}
{\bf D} {\bf U}^H$ where ${\bf D}$ is the $N_T \times N_T$
diagonal matrix having the eigenvalues in (\ref{eq:rc_eigs1}) as
its diagonal entries, and ${\bf U}$ is a unitary matrix.
Substituting this into the right hand side of
(\ref{eq:rc_capacity_RicianGeneral4}), we have
\begin{eqnarray}
C_{N_T,N_R} &\leq& {\underset{ {\underset{ {\bf Q} \ \geq \
0}{{\textrm{tr}} \ {\bf Q} \ \leq \ P}} }{\max}} \ \ \log \det
\left(N_R \tilde{ {\bf Q}}  {\bf D} + {\bf I}_{N_T} \right) ,
\label{eq:rc_capacity_RicianGeneral5}
\end{eqnarray}
where we have let ${\bf U}^H {\bf Q} {\bf U} = \tilde{ {\bf Q}}$.
Now it is easy to see that the right-hand side of
(\ref{eq:rc_capacity_RicianGeneral5}) is maximized by a diagonal
$\tilde{ {\bf Q}}$ and the diagonal entries are again given by the
well-known water-filling\index{water-filling} algorithm. Indeed,
one can show that the maximizing diagonal matrix $\tilde{ {\bf
Q}}$ is such that,
\begin{eqnarray}
\tilde{ {\bf Q}}_{i,i} &=& \left\{ \begin{array}{cc} \min \left\{
\frac{P}{N_T},  \frac{\kappa(1+ \kappa)}{N_R (1 + N_T \kappa)}
\right\} N_T + \left[\frac{P}{N_T} - \frac{\kappa(1+ \kappa)}{N_R
(1 + N_T \kappa)} \right]^{+} \ & \
\textrm{if $i = 1$} \\
\left[\frac{P}{N_T} - \frac{\kappa(1+ \kappa)}{N_R (1 + N_T
\kappa)} \right]^{+} \ & \ \textrm{if $i = 2, \ldots , N_T$}
\end{array} \right. \ \ \ {\textrm{for}} \ \ \ 0 \leq \kappa <
\infty.  \nonumber
\end{eqnarray}
where $\left[ x \right]^{+} = \max \{0,x \}$.

Thus, for $0 \leq \kappa < \infty$, the capacity of a multiple
antenna system in Rician fading, subjected to the average transmit
power constraint $P$, with perfect CSI at the receiver and
knowledge only of $\kappa$ at the transmitter is upper bounded as
\begin{eqnarray}
C_{N_T,N_R} &\leq& \log \left(1 + \frac{N_R \left(1+ N_T \kappa
\right)}{1 + \kappa} \left( \min \left\{ \frac{P}{N_T},
\frac{\kappa(1+ \kappa)}{N_R (1 + N_T \kappa)} \right\} N_T +
\left[\frac{P}{N_T} - \frac{\kappa(1+ \kappa)}{N_R (1 + N_T
\kappa)} \right]^{+} \right) \right) \nonumber \\
& & ~~~~~~ + (N_T-1) \log \left(1 + \frac{N_R}{1 + \kappa}
\left[\frac{P}{N_T} - \frac{\kappa(1+ \kappa)}{N_R (1 + N_T
\kappa)} \right]^{+} \right) .
\label{eq:rc_capacity_RicianGeneral6}
\end{eqnarray}
(For $\kappa \longrightarrow \infty$, the exact capacity is given
by (\ref{eq:rc_capacity_RicianGeneral11}).)

Next, we will illustrate the above bound for some special cases.

\subsection*{Case 1: $\kappa = 0$}

It is easily seen that for $\kappa = 0$, which is the Rayleigh
fading\index{Rayleigh fading} case, the above bound reduces to
\begin{eqnarray}
C_{N_T,N_R} &\leq& N_T \log \left(1 + \frac{N_R}{N_T}  P \right) \
\ \ {\textrm{for}} \ \ \ \kappa \ = \ 0.
\label{eq:rc_capacity_RicianGeneral7}
\end{eqnarray}

From (\ref{eq:rc_capacity_RicianGeneral7}), we see that when $N_R
= N_T$ the capacity upper bound is a linear function of $N_T$. In
fact, it was shown in \cite{Telatar1} that in this case the
capacity can be approximated by a linear function of $N_T$
asymptotically for large numbers of antennas.

In addition, if $N_T=1$ in (\ref{eq:rc_capacity_RicianGeneral7}),
then
\begin{eqnarray}
C_{N_T,N_R} &\leq& \log \left(1 + N_R  P \right)  \ \ \
{\textrm{for}} \ \ \ \kappa \ = \ 0 \ \ \ {\textrm{and}} \ \ \
N_T=1. \label{eq:rc_capacity_RicianGeneral7b}
\end{eqnarray}

In fact, it was shown in \cite{Telatar1} that the capacity of the
Rayleigh fading\index{Rayleigh fading} channel in this case is
asymptotic to $\log (1 + N_R P)$ for large $N_R$.

\subsection*{Case 2: $N_T = 1$}

When $N_T = 1$, the capacity upper bound in
(\ref{eq:rc_capacity_RicianGeneral6}) becomes
\begin{eqnarray}
C_{1,N_R} &\leq& \log \left(1 + N_R  P \right) \ \ \
{\textrm{for}} \ \ \ \kappa \ \geq \ 0 .
\label{eq:rc_capacity_RicianGeneral8}
\end{eqnarray}
From (\ref{eq:rc_capacity_RicianGeneral8}) we see that the bound
(\ref{eq:rc_capacity_RicianGeneral7b}) is in fact valid not only
for the Rayleigh channel but for any Rician channel with an
arbitrary value of $\kappa$, in this case of $N_T = 1$.

\subsection*{Case 3: $N_R = 1$}

The capacity of the Rician channel is bounded in this case as
\begin{eqnarray}
C_{N_T,1} &\leq& \log \left(1 + \frac{1+ N_T \kappa}{1 + \kappa}
\left( \min \left\{ \frac{P}{N_T}, \frac{\kappa(1+ \kappa)}{1 +
N_T \kappa} \right\} N_T + \left[\frac{P}{N_T} - \frac{\kappa(1+
\kappa)}{1 + N_T
\kappa} \right]^{+} \right) \right) \nonumber \\
& & ~~~~~~ + (N_T-1) \log \left(1 + \frac{1}{1 + \kappa}
\left[\frac{P}{N_T} - \frac{\kappa(1+ \kappa)}{1 + N_T \kappa}
\right]^{+} \right) 
\label{eq:rc_capacity_RicianGeneral9} .
\end{eqnarray}

\subsection*{Case 4: $N_R = N_T = n$}

In this case, the capacity upper bound of
(\ref{eq:rc_capacity_RicianGeneral6}) reduces to
\begin{eqnarray}
C_{n,n} &\leq& \log \left(1 + \frac{1+ n \kappa}{1 + \kappa}
\left( \min \left\{ P, \frac{\kappa(1+ \kappa)}{1 + n \kappa}
\right\} n + \left[P - \frac{\kappa(1+ \kappa)}{1 + n
\kappa} \right]^{+} \right) \right) \nonumber \\
& & ~~~~~~ + (n-1) \log \left(1 + \frac{1}{1 + \kappa} \left[P -
\frac{\kappa(1+ \kappa)}{1 + n \kappa} \right]^{+} \right)
\nonumber 
\end{eqnarray}

Note that the exact capacity as $\kappa \longrightarrow \infty$
equals $\log(1 + n^2 P) $ in this case.

\section{Capacity of the Rician Channel When the Transmitter Does
Not Know The Fading Distribution} \label{sec:rc_RiceCap2}

In this section, we investigate the capacity of the
average-power-constrained Rician channel when the receiver (but
not the transmitter) has perfect CSI, and the transmitter does not
know the fading distribution. Of course, this capacity provides a
lower bound for the capacity in the situation where the
transmitter does know the fading distribution. Recall from the
preceding section that the optimal transmitted signal distribution
when the transmitter does not know the fading distribution is a
circularly symmetric complex Gaussian distribution with $\emph{E}
\{ {\bf x} {\bf x}^H \} = \frac{P}{N_T} {\bf I}_{N_T}$. Since this
distribution is also optimal for the average-power-constrained
Rayleigh channel, evaluation of the capacity with this signal
distribution in the Rician channel also quantifies the capacity
variation of a system designed to be optimal for Rayleigh fading
but operating in a Rician channel (i.e. a channel with a
line-of-sight component).

Applying this signal distribution, from
(\ref{eq:rc_capacity_RicianGeneral}) the capacity of the Rician
channel (arbitrary $\kappa$) with no transmitter knowledge of
$\kappa$ is given by
\begin{eqnarray}
C_{N_T,N_R} &=& \emph{E}_{{\bf W}} \left\{ \log \det \left(
\frac{P}{N_T} {\bf W} +  {\bf I}_{n} \right) \right\} ,
\label{eq:rc_capacity_Rician3}
\end{eqnarray}
where the expectation is with respect to ${\bf W} \sim {\cal W}_m
\left(n, {\bm \Sigma}, {\bm \Sigma}^{-1}{\bf M} {\bf M}^H \right)$
with pdf (\ref{eq:rc_nonCentrlWishart2}).

Since ${\bf W}$ in (\ref{eq:rc_capacity_Rician3}) is an $m \times
m$ Hermitian matrix, if we denote its (non-negative) eigenvalues
by $\lambda_1,\lambda_2, \ldots,\lambda_m$, then we have $\det
\left( \frac{P}{N_T} {\bf W} + {\bf I}_m \right) = \prod_{i=1}^m
\left( 1 + \frac{P}{N_T} \lambda_i \right)$. Hence, the capacity
in (\ref{eq:rc_capacity_Rician3}) can be given in terms of the
eigenvalue distribution of the non-central Wishart distributed
matrix ${\bf W}$ as
\begin{eqnarray}
C_{N_T,N_R} &=& \emph{E}_{\lambda_1,\ldots,\lambda_m} \left\{ \log
\prod_{i=1}^m \left( 1 + \frac{P}{N_T} \lambda_i \right) \right\}
, \label{eq:rc_capacity_Rician4} \\
&=& m \emph{E}_{\lambda} \left\{ \log \left( 1 + \frac{P}{N_T}
\lambda \right) \right\} \ , \label{eq:rc_capacity_Rician6}
\end{eqnarray}
where in (\ref{eq:rc_capacity_Rician6}) we have taken $\lambda$ to
be any un-ordered eigenvalue of the non-central Wishart
distributed random matrix\index{Wishart matrix} ${\bf W}$.

\subsection{Special Case $1$: Capacity in the Limit of Large $N_T$}
\label{sec:RicianCapLarge_t}

Before attempting to evaluate the capacity exactly, it is
instructive to investigate its behavior in the limit as the number
of transmit antennas increases without bound. This will also allow
us to compare the asymptotic capacity in this Rician case with
that of the Rayleigh case given in \cite{Telatar1}, thereby
illustrating the effect of the non-zero mean of fading
coefficients on the capacity.

Note that for a fixed $N_R$, the elements of the matrix ${\bf
H}{\bf H}^H$ are the sums of $N_T$ iid random variables with
finite moments, and thus by the strong law of large numbers (SLLN)
we have, almost surely, $\lim_{N_T \longrightarrow \infty}
\frac{1}{N_T} {\bf H}{\bf H}^H = {\bm \Upsilon}$ where the matrix
${\bm \Upsilon}$ is defined in (\ref{eq:rc_ExpectMat1}) (it is now
taken to be an $m \times m$ square matrix). Thus, for a fixed
number of receive antennas, when the number of transmit antennas
becomes very large, the capacity of this channel is given by
\begin{eqnarray}
C_{\infty,N_R} &=& \lim_{N_T \longrightarrow \infty} \log \det
\left( {\bf I}_{N_R} + \frac{P}{N_T} {\bf H}{\bf H}^H \right)
\ = \ \log \det \left( {\bf I}_{N_R} + P {\bm \Upsilon} \right) , \nonumber \\
&=& (N_R-1) \log \left[ 1 + \frac{P}{1+\kappa} \right]  + \log
\left[1 + (N_R \kappa + 1) \frac{P}{1+\kappa} \right] .
\label{eq:rc_capacity_rician_asymp2}
\end{eqnarray}

The following two cases illustrate the dependence of the above
asymptotic capacity on the Rician factor\index{Rician factor}
$\kappa$:
\begin{eqnarray}
\lim_{\kappa \longrightarrow 0}  C_{\infty,N_R} &=& N_R \log
\left(1 +P \right)  \label{eq:rc_capacity_rician_asymp3}
\end{eqnarray}
and
\begin{eqnarray}
\lim_{\kappa \longrightarrow \infty} C_{\infty,N_R} &=& \log
\left( 1 + N_R P \right) . \label{eq:rc_capacity_rician_asymp4}
\end{eqnarray}

Note that (\ref{eq:rc_capacity_rician_asymp3}) is the asymptotic
capacity of the Rayleigh channel given in \cite{Telatar1}. It is
easily seen that $C_{\infty,N_R}$ in
(\ref{eq:rc_capacity_rician_asymp2}) is a monotonically decreasing
function of $\kappa$ for $\kappa > 0$ and $N_R >1$ and is constant
for $N_R = 1$. Thus, we observe that when $N_T$ is very large and
$N_R>1$, increasing the determinism of the channel lowers the
capacity if the transmitter is not aware of this increase, and
moreover the Rician fading environment will degrade the capacity
of a system that is designed to achieve the Rayleigh channel
capacity. Of course, this does not necessarily mean that the
capacity of the Rician fading channel is less than that of the
Rayleigh fading channel when the transmitter knows $\kappa$. In
fact, from (\ref{eq:rc_capacity_RicianGeneral11}) we may recall
that in the deterministic case, which is the limiting case of
Rician fading with $\kappa \longrightarrow \infty$, the capacity
of the channel, as achieved by the water-filling algorithm, is
known to be $\log(1 + N_R N_T P)$, for any $N_R$ and $N_T$.
Similarly, it is reasonable to expect that the capacity of a
multiple antenna system in Rician fading, with an arbitrary
$\kappa$, to be greater than that in a Rayleigh fading channel
when transmitter knows the value of $\kappa$.

\begin{figure}[hbt!]
\centering
\includegraphics[width=5in]{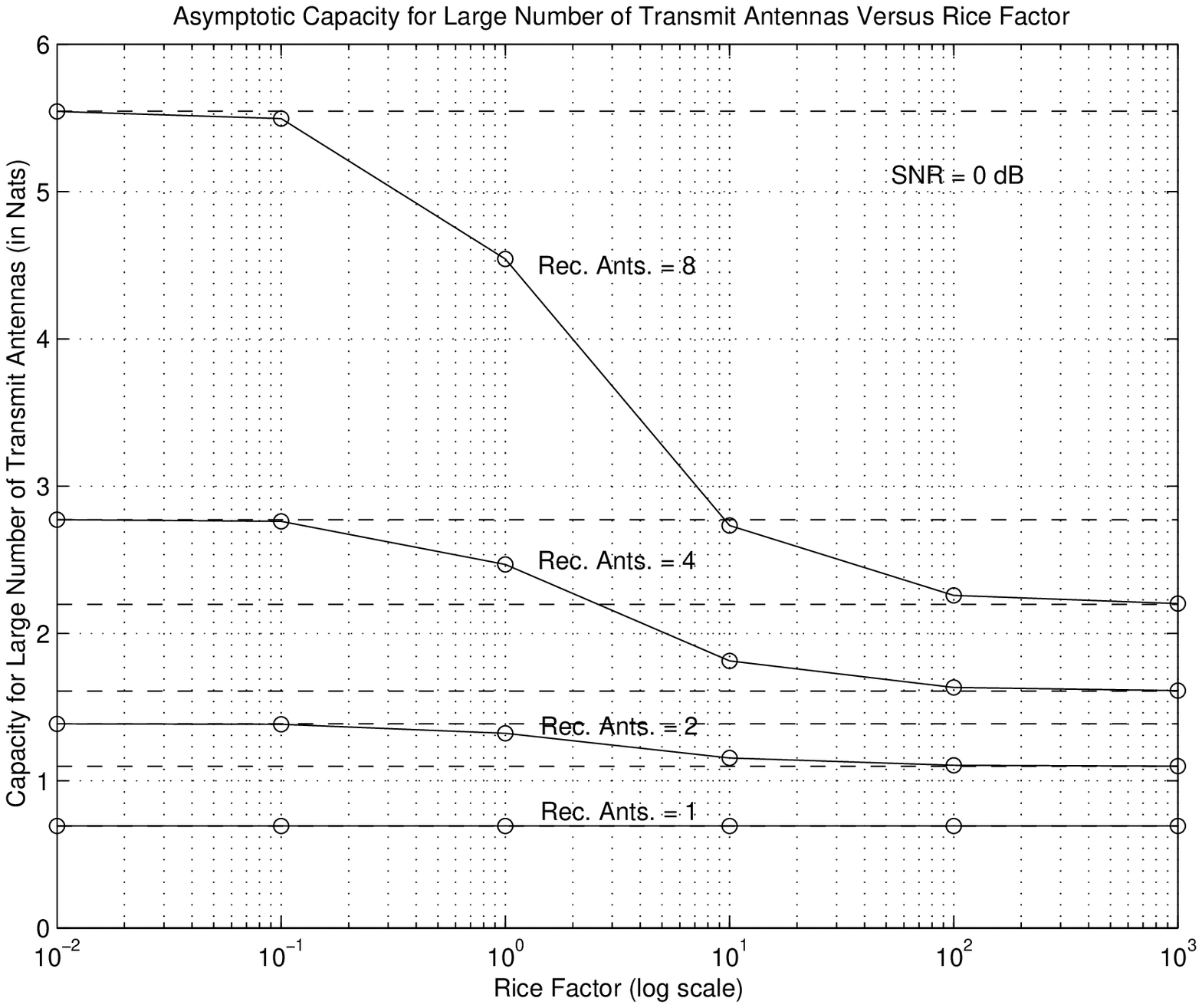}
\caption{Asymptotic Capacity for Large Number of Transmit Antennas
Versus Rician Factor. SNR $ = 0 \ dB$. } \label{fig:CapAsymp1_1}
\end{figure}

\begin{figure}[hbt!]
\centering
\includegraphics[width=5in]{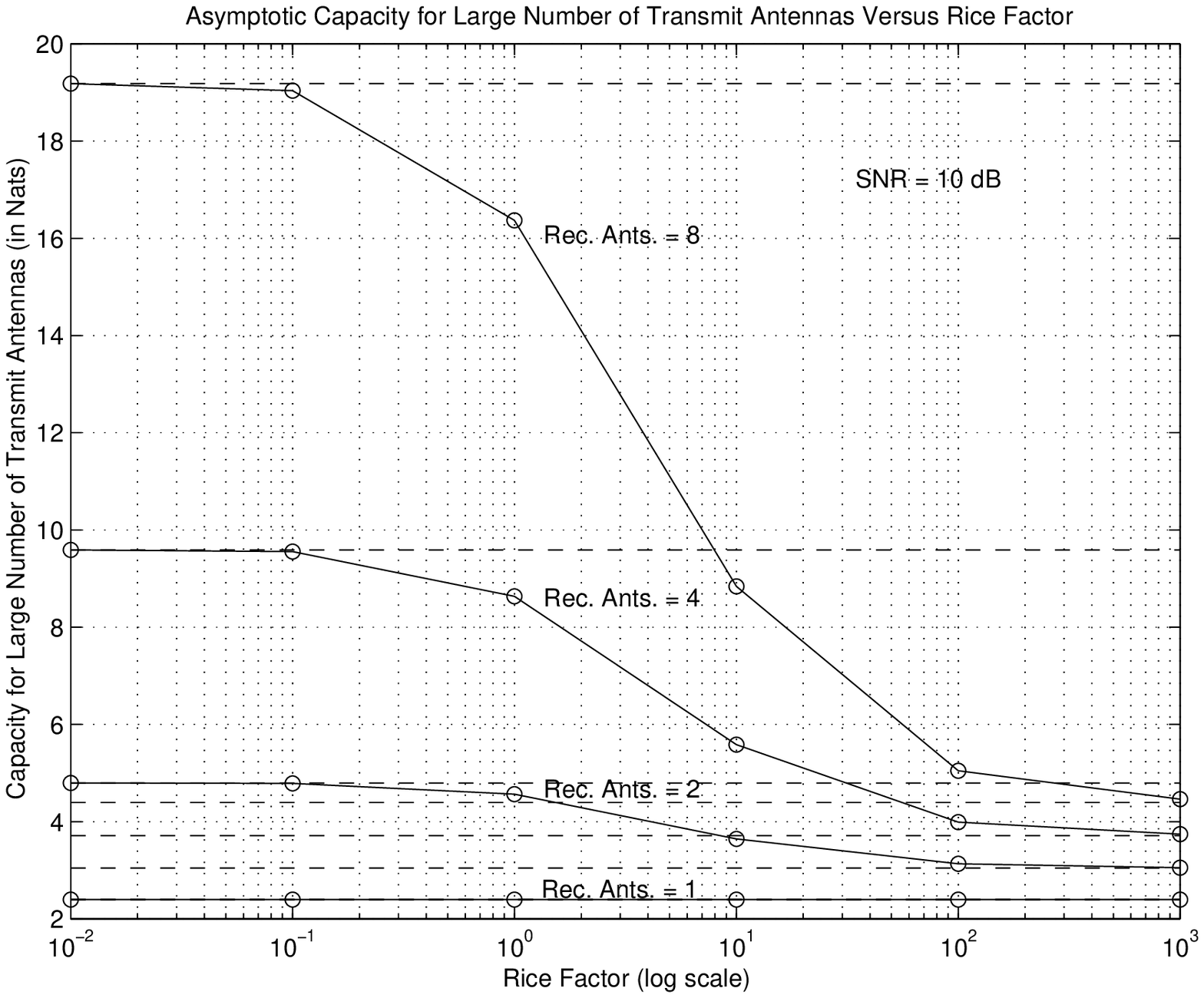}
\caption{Asymptotic Capacity for Large Number of Transmit Antennas
Versus Rician Factor. SNR $ = 10 \ dB$. } \label{fig:CapAsymp1_2}
\end{figure}

Figures \ref{fig:CapAsymp1_1} and \ref{fig:CapAsymp1_2} show the
dependence of the asymptotic capacity
(\ref{eq:rc_capacity_rician_asymp2}) on the Rician
factor\index{Rician factor} $\kappa$ for $P = 0$ dB and $P = 10$
dB, respectively. These graphs illustrate our conclusion about the
asymptotic capacity degradation of the Rician channel with
increasing $\kappa$ when the transmitter has no knowledge of
$\kappa$.

\begin{figure}[hbt!]
\centering
\includegraphics[width=5in]{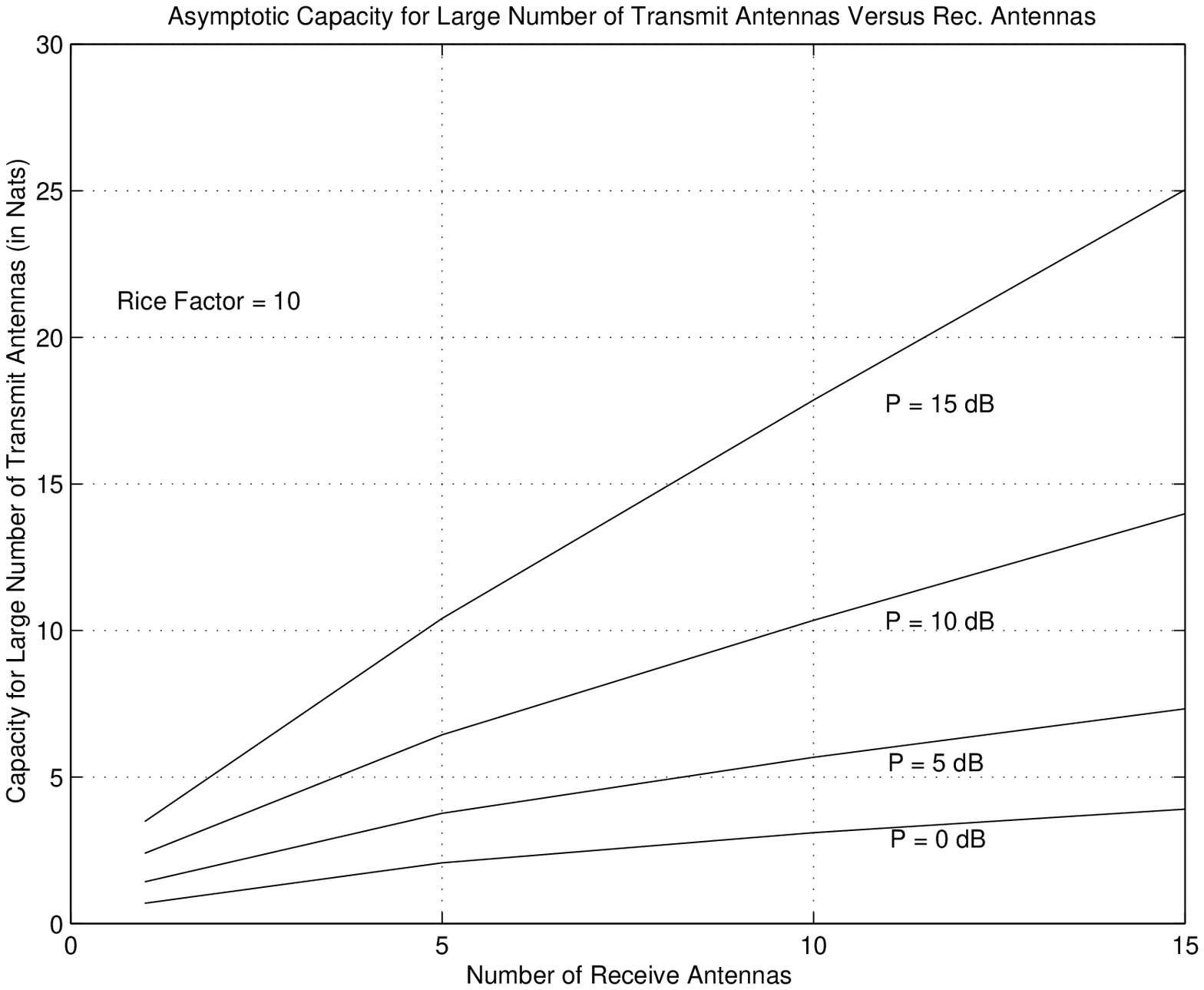}
\caption{Asymptotic Capacity for Large Number of Transmit Antennas
Versus Number of Receiver Antennas.  $ \kappa = 10$. }
\label{fig:CapAsymp2_linear2}
\end{figure}

In Fig. \ref{fig:CapAsymp2_linear2} we show the asymptotic
capacity (\ref{eq:rc_capacity_rician_asymp2}) versus the number of
receiver antennas $N_R$ for $ \kappa = 10$. Figure
\ref{fig:CapAsymp2_linear2} shows the almost linear dependence of
this capacity on the number of receiver antennas (which is the
smaller of $N_T$ and $N_R$ in this case), similarly to the
previously established linear dependence for the Rayleigh fading
environment (\cite{Foschini1,Telatar1}).

\subsection{Special Case $2$ : $\min \{ N_R, N_T \} = 1$}

In this section we will evaluate the exact capacity of the Rician
fading channel (subject to the signal choice assumed throughout
this section) in the special case $m = \min \{ N_R, N_T \} = 1$.
In this special case (\ref{eq:rc_Cov1}) reduces  to a scalar ${\bm
\Sigma} = 2 \sigma^2 $, and thus the pdf of ${\bf W}$ in
(\ref{eq:rc_nonCentrlWishart2}) (which is a scalar) can be written
as
\begin{eqnarray}
f_{{\bf W}}(W) &=& e^{-\kappa n } \frac{1}{ (2 \sigma^2)^n}
e^{-\frac{W}{2 \sigma^2}} |W|^{n-1} \left(  \sqrt{n \kappa (1 +
\kappa) W} \right)^{-(n-1)} I_{n-1} \left(2 \sqrt{n \kappa (1 +
\kappa) W} \right) , \label{eq:rc_nonCentrlWishartSpecial3}
\end{eqnarray}
where we have used the fact that $\tilde{\Gamma}_1(a) =   \Gamma
(a)$ and (\ref{eq:rc_BesselFuncOfMatrixSpecial1}) from the
Appendix.

From (\ref{eq:rc_capacity_Rician3}) the capacity in this special
case is
\begin{eqnarray}
C_{N_T,N_R} &=& \int_0^{\infty} \log \left( 1 + \frac{P}{N_T} W
\right) f_{{\bf W}}(W) dW , \label{eq:rc_capacity_RicianSpecial1}
\end{eqnarray}
where $f_{{\bf W}}(W)$ is given in
(\ref{eq:rc_nonCentrlWishartSpecial3}) above. As noted by Telatar
in \cite{Telatar1} for the Rayleigh fading\index{Rayleigh fading}
channel, from (\ref{eq:rc_capacity_RicianSpecial1}) we observe
that the capacity is not symmetric in $N_R$ and $N_T$ also in the
Rician case. Thus, we have two cases to consider as below.

\subsubsection{Case $1$: $N_R \geq N_T = 1$}
From (\ref{eq:rc_nonCentrlWishartSpecial3}) and
(\ref{eq:rc_capacity_RicianSpecial1}), the capacity of the Rician
fading channel when the transmitter does not know $\kappa$ in this
case is,
\begin{eqnarray}
C_{1,N_R} &=& \frac{1}{\Gamma(N_R)} \int_0^{\infty} \log \left( 1
+ P W \right) W^{N_R-1} e^{-W} \psi (W,N_R) dW ,
\label{eq:rc_capacity_RicianXmitr1}
\end{eqnarray}
where we have introduced the function
\begin{eqnarray}
\psi (W,n) &=& \frac{\Gamma(n)}{[n \kappa (1+
\kappa)]^{\frac{1}{2}(n-1)}} \left( \frac{1 + \kappa}{e^{\kappa}}
\right)^n e^{-\kappa W } W^{-\frac{1}{2}(n-1)} I_{n-1} \left(2
\sqrt{n \kappa (1 + \kappa ) W} \right) . \label{eq:rc_dummyFunc1}
\end{eqnarray}

\begin{figure}[hbt!]
\centering
\includegraphics[width=5in]{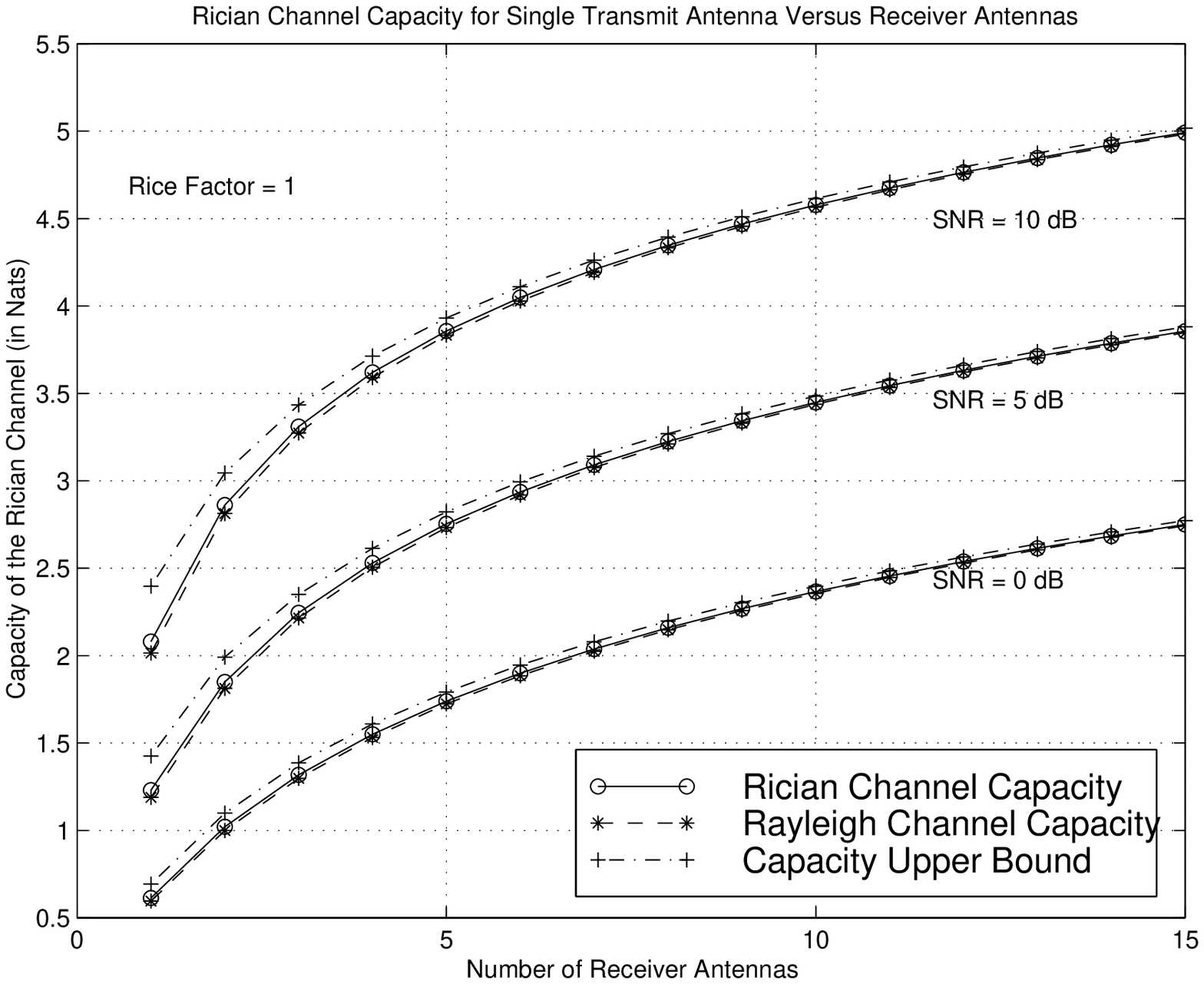}
\caption{Rician Channel Capacity for Single Transmit Antenna
Versus Receiver Antennas.  $ \kappa = 1$. }
\label{fig:Cap_n1_t1_kappa1}
\end{figure}

\begin{figure}[hbt!]
\centering
\includegraphics[width=5in]{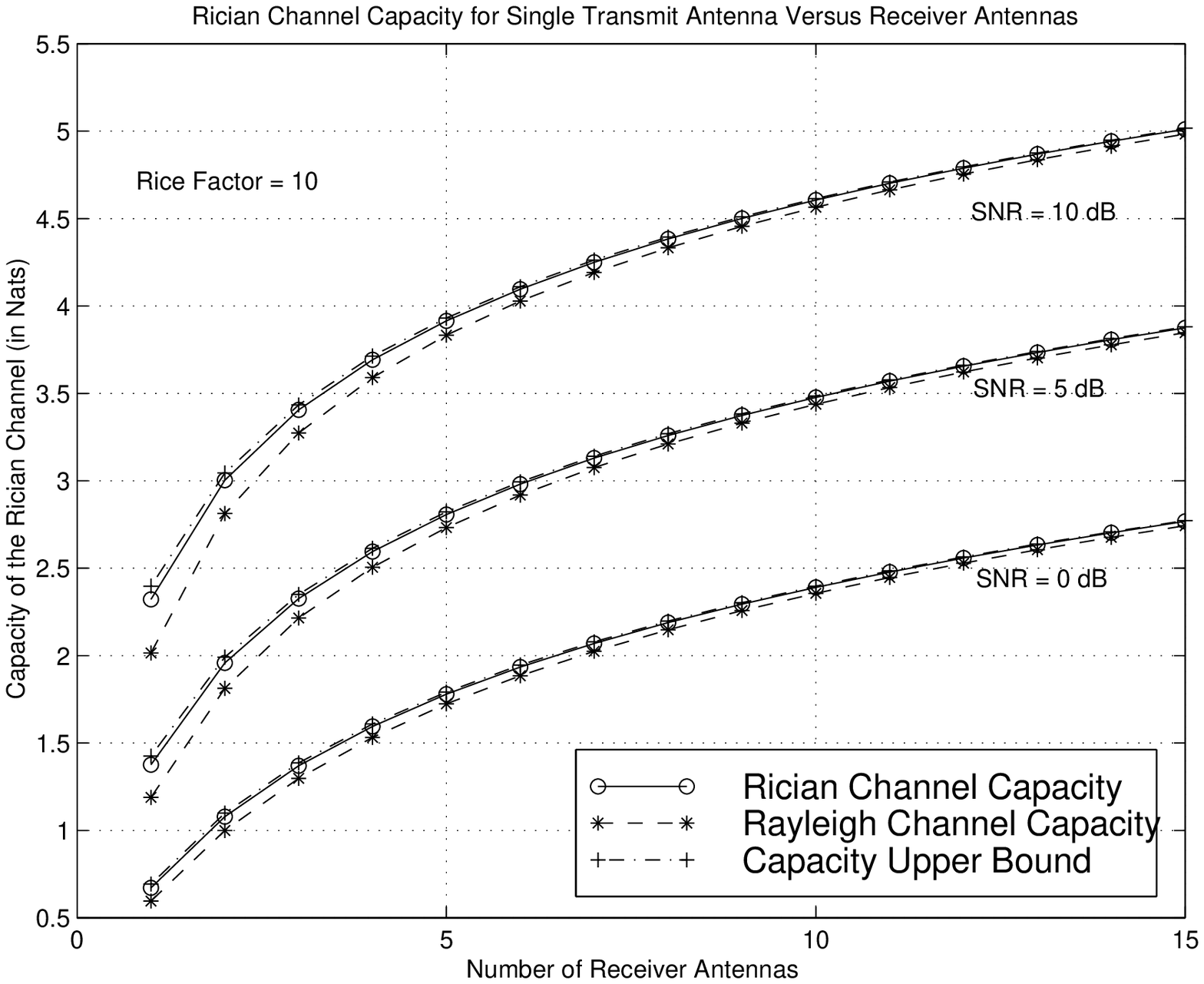}
\caption{Rician Channel Capacity for Single Transmit Antenna
Versus Receiver Antennas.  $ \kappa = 10$. }
\label{fig:Cap_n1_t1_kappa10}
\end{figure}

Figures \ref{fig:Cap_n1_t1_kappa1} and \ref{fig:Cap_n1_t1_kappa10}
show the capacity of the Rician fading channel when the
transmitter does not know $\kappa$ in this special case of $N_T =
1$, against the number of receiver antennas for $\kappa = 1$ and
$\kappa = 10$, respectively. Included on the same graphs are the
corresponding capacity curves for the Rayleigh fading channel
($\kappa = 0$). We observe that the capacity of the Rician channel
is greater than that of the Rayleigh channel and the capacity gap
increases with increasing values of $\kappa$. We can also observe
from Figs. \ref{fig:Cap_n1_t1_kappa1} and
\ref{fig:Cap_n1_t1_kappa10} that the capacity gap is prominent for
smaller numbers of receiver antennas and, as $N_R \longrightarrow
\infty$, the two capacities converge to the same value. In Figs.
\ref{fig:Cap_n1_t1_kappa1} and \ref{fig:Cap_n1_t1_kappa10} we have
also shown the capacity upper bound for this system given by
(\ref{eq:rc_capacity_RicianGeneral8}). It is clear for these
figures that the upper bound (\ref{eq:rc_capacity_RicianGeneral8})
is very tight in this case. Indeed, since $N_T=1$, in this case
the optimal signal covariance that satisfies the average power
constraint is ${\bf Q} = Q = E \{ x^2 \} = P$.

\subsubsection{Case $2$: $N_T \geq N_R = 1$} \label{sec:rc_RiceCap2_t1}
The capacity of the Rician fading channel when the transmitter
does not know $\kappa$ in this case is
\begin{eqnarray}
C_{N_T,1} &=& \frac{1}{\Gamma(N_T)} \int_0^{\infty} \log \left( 1
+ \frac{P}{N_T} W \right) W^{N_T-1} e^{-W} \psi (W,N_T) dW ,
\label{eq:rc_capacity_RicianRcvr1}
\end{eqnarray}
where $\psi (W,N_T)$ is given by (\ref{eq:rc_dummyFunc1}).

\begin{figure}[hbt!]
\centering
\includegraphics[width=5in]{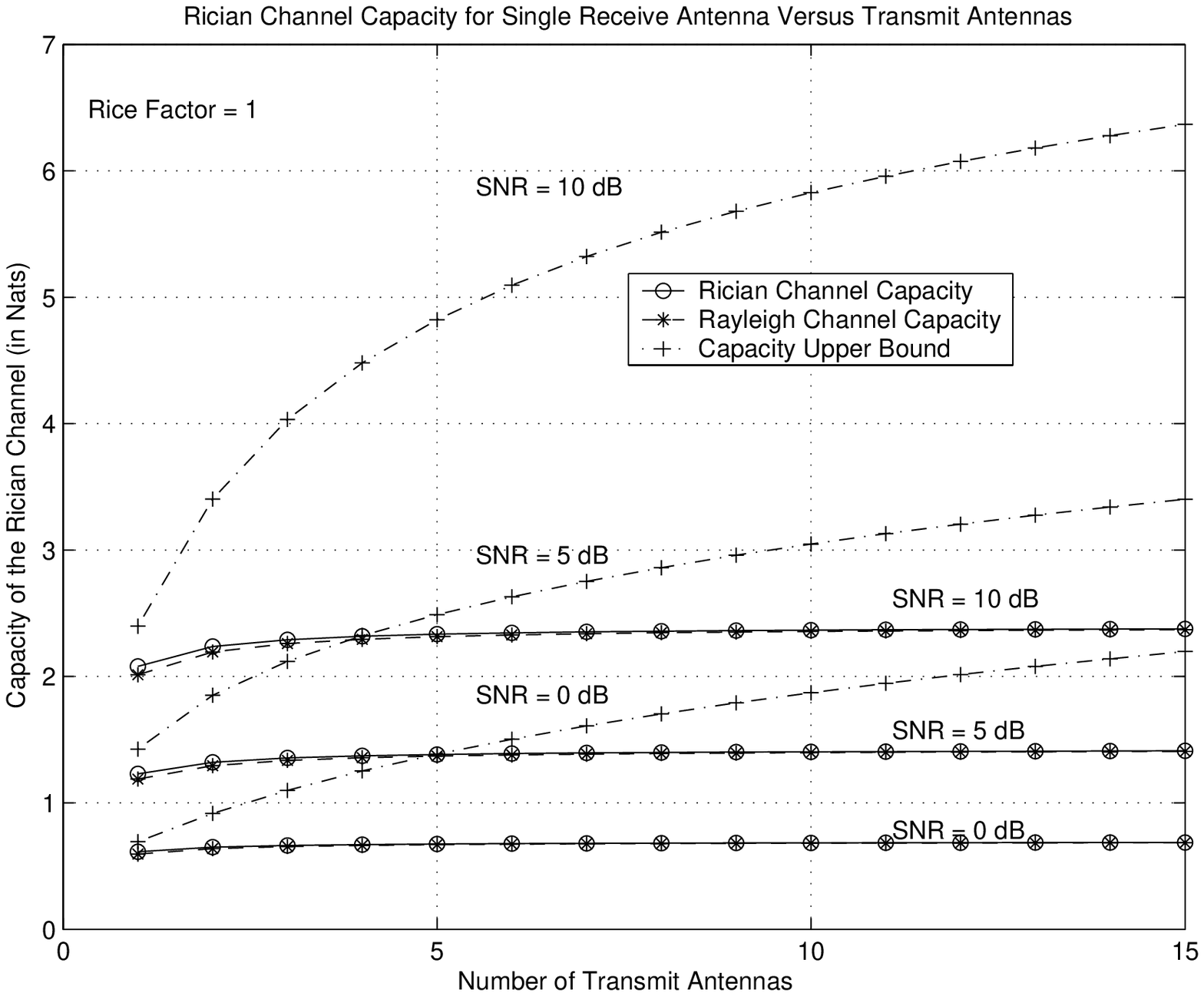}
\caption{Rician Channel Capacity for Single Receiver Antenna
Versus Transmit Antennas.  $ \kappa = 1$. }
\label{fig:Cap_n1_r1_kappa1}
\end{figure}

\begin{figure}[hbt!]
\centering
\includegraphics[width=5in]{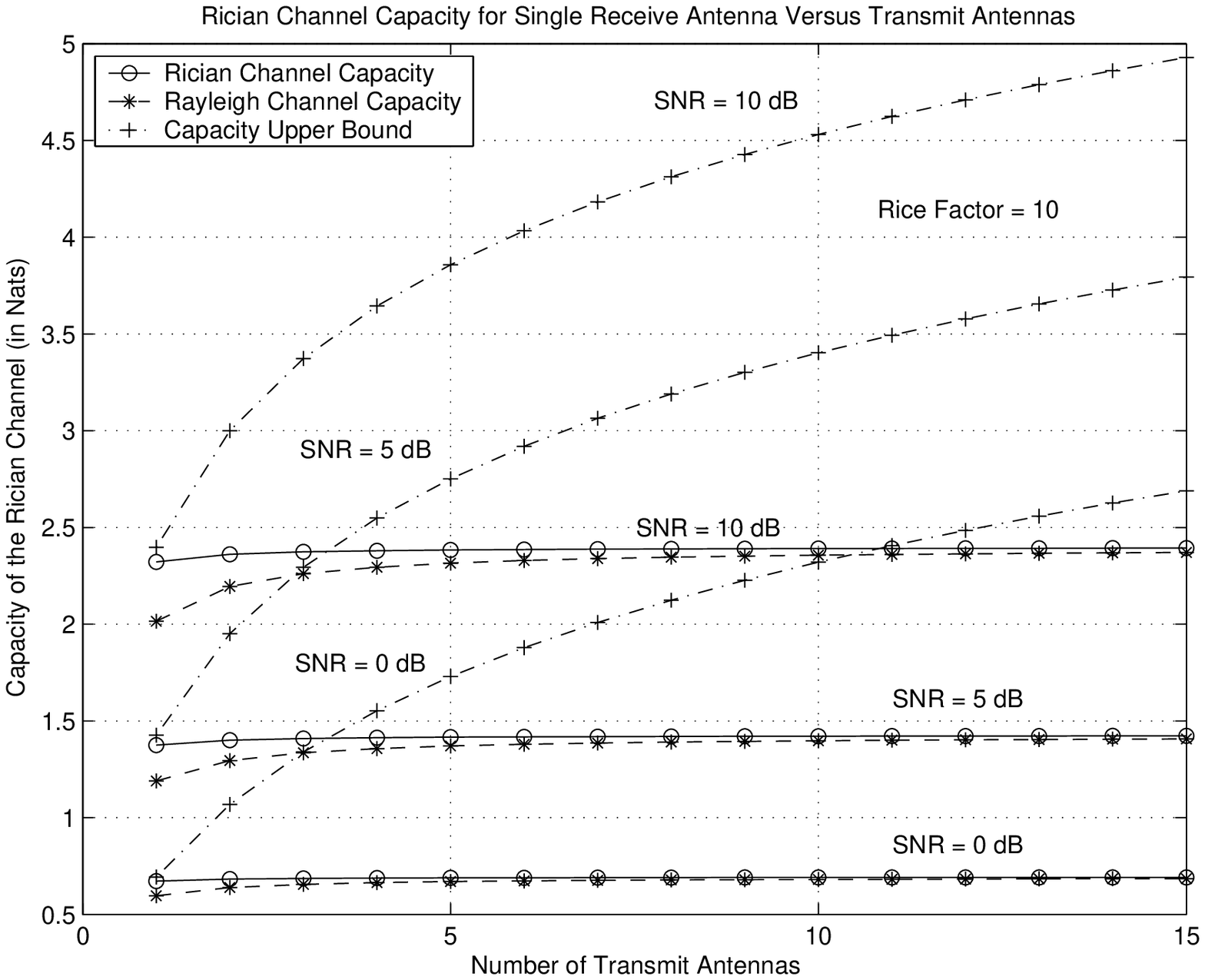}
\caption{Rician Channel Capacity for Single Receiver Antenna
Versus Transmit Antennas.  $ \kappa = 10$. }
\label{fig:Cap_n1_r1_kappa10}
\end{figure}

Figures \ref{fig:Cap_n1_r1_kappa1} and \ref{fig:Cap_n1_r1_kappa10}
plot the capacity of the Rician fading channel in this special
case of $N_R = 1$, against the number of transmit antennas for
$\kappa = 1$ and $\kappa = 10$, respectively. As before, included
on the same graphs are the corresponding capacity curves for the
Rayleigh fading channel (i.e. the $\kappa = 0$ case). Again, we
observe that the capacity of the Rician channel is greater than
the capacity of the Rayleigh channel and the capacity gap
increases with increasing values of $\kappa$ before finally
converging to the same value for large $N_T$. Particularly, from
Fig. \ref{fig:Cap_n1_r1_kappa10} we note that for a smaller number
of transmit antennas the capacity gap is significant.

Shown also on Figs. \ref{fig:Cap_n1_r1_kappa1} and
\ref{fig:Cap_n1_r1_kappa10} is the capacity upper bound for the
$N_R=1$ case given by (\ref{eq:rc_capacity_RicianGeneral9}). From
these figures we observe that, unlike in the case of $N_T=1$, the
upper bound is very loose for the case of $N_R=1$. However, recall
that we are using a particular input signal distribution which is
not necessarily the capacity achieving distribution for this
particular channel under the assumed conditions on CSI and fading
statistics. Rather, we were assuming a signal distribution that is
only optimal for the Rayleigh fading channel or for a system in
which transmitter does not know the value of $\kappa$. Thus Figs.
\ref{fig:Cap_n1_r1_kappa1} and \ref{fig:Cap_n1_r1_kappa10} suggest
that scaled identity matrix might not be the form of the
covariance matrix of the capacity achieving input signal
distribution for a multiple antenna Rician channel when the
transmitter knows $\kappa$, and with better signal choices that
exploit the {\it Rician-ness} inherent in the fading distribution,
we may be able to obtain higher capacities. In Section
\ref{sec:rc_newSigs} we propose a new choice for the covariance
matrix which offers much higher capacity than that achieved by the
scaled identity matrix.

Using a series representation of the modified Bessel function, it
is straightforward to show that $\lim_{\kappa \longrightarrow 0}
\psi (W,m) = 1 $. Thus, in the  limit ${\kappa \longrightarrow
0}$, (\ref{eq:rc_capacity_RicianXmitr1}) and
(\ref{eq:rc_capacity_RicianRcvr1}) reduce to the corresponding
capacity expressions for the Rayleigh channel given in
\cite{Telatar1}, as one would have expected.

\subsection{General Capacity Expression for the Rician Channel}

In order to compute the capacity of the Rician channel for an
arbitrary number of transmit/receiver antennas, as given in
(\ref{eq:rc_capacity_Rician6}), we need to find the latent root
distribution of the non-central Wishart distributed
matrix\index{Wishart distribution!non-central} ${\bf W}$. This
latent root distribution has been studied previously
(\cite{Constantine1,James3,James6,James1}) and, in particular, we
have the following result from \cite{James1}.
\begin{theorem}
If ${\bf W}$ has the non-central Wishart
distribution\index{Wishart distribution!non-central} given in
(\ref{eq:rc_nonCentrlWishart2}), then the pdf of the latent roots
$\hat{ {\bm \Lambda}} = {\textrm{diag}}(\hat{\lambda}_1, \ldots,
\hat{\lambda}_m)$ of $|{\bf W} - \hat{\lambda} {\bm \Sigma}| = 0$
depends only on the latent roots $\hat{{\bm \Omega}} =
{\textrm{diag}}(\hat{\omega}_1, \ldots, \hat{\omega}_m)$ of $|{\bf
M} {\bf M}^H - \hat{\omega} {\bm \Sigma}| = 0$, and is given by
\begin{eqnarray}
f_{\hat{ {\bm \Lambda}}}(\hat{{\bm \Lambda}}) &=& e^{-{\textrm tr}
\hat{{\bm \Omega}}} {}_0\tilde{F}_1 (n;\hat{{\bm \Omega}},\hat{
{\bf \Lambda}}) \frac{\pi^{m(m-1)}}{\tilde{\Gamma}_{m}(n)
\tilde{\Gamma}_{m}(m)} e^{- {\textrm tr} \hat{ {\bf \Lambda}}}
|\hat{ {\bf \Lambda}}|^{n-m} \prod_{i<j}^m (\hat{\lambda}_i -
\hat{\lambda}_j)^2 . \label{eq:rc_nonCentrlWishartLatentRoots1}
\end{eqnarray}
\end{theorem}

Due to the scaled identity matrix structure of the covariance
matrix ${\bm \Sigma}$ in our case (see (\ref{eq:rc_Cov1})), it is
easily seen that the latent root distribution ${\bm \Lambda}$ of
the matrix ${\bf W}$, as required in
(\ref{eq:rc_capacity_Rician4}), can be obtained from the
distribution given in (\ref{eq:rc_nonCentrlWishartLatentRoots1})
by noting that $ 2 \sigma^2 \hat{\lambda}_i = \lambda_i$ and $ 2
\sigma^2 \hat{\omega}_i = \omega_i$ for $i = 1, \ldots m$, and
thus $2 \sigma^2 \hat{ {\bf \Lambda}} = {\bf \Lambda}$ and $ 2
\sigma^2 \hat{{\bm \Omega}} = {\bm \Omega}$, where we have denoted
the latent roots matrix of ${\bf M} {\bf M}^H$ by ${\bm \Omega}$.
Hence, by applying this change of variables to
(\ref{eq:rc_nonCentrlWishartLatentRoots1}), we get the required
eigenvalue distribution of the matrix ${\bf W}$ as
\begin{eqnarray}
f_{{\bf \Lambda}}(\Lambda) = (1+ \kappa)^n e^{-\frac{1}{2
\sigma^2} {\textrm tr} \Omega} {}_0\tilde{F}_1 (m;\frac{1}{2
\sigma^2} \Omega, \frac{1}{2 \sigma^2}{\bf \Lambda})
\frac{\pi^{n(n-1)}}{\tilde{\Gamma}_{n}(m) \tilde{\Gamma}_{n}(n)}
 e^{- tr \frac{1}{2 \sigma^2} {\bf \Lambda}} |\frac{1}{2
\sigma^2} {\bf \Lambda}|^{m-n} \prod_{i<j}^n \left(
\frac{\lambda_i}{2 \sigma^2} - \frac{\lambda_j}{2 \sigma^2}
\right)^2 . \label{eq:rc_nonCentrlWishartLatentRoots2}
\end{eqnarray}

From the definition of ${\bf M}$ in (\ref{eq:rc_Mean1}) we see
that ${\bf M} {\bf M}^H = \mu^2 {\bm \Psi}_{m,n} {\bm
\Psi}_{m,n}^H = n {\mu}^2 {\bm \Psi}_{m}$ where ${\bm \Psi}_{m}$
denotes the $m \times m$ matrix of all ones. By decomposing ${\bm
\Psi}_{m}$ as ${\bm \Psi}_{m} = m {\bf u} {\bf u}^H$ where ${\bf
u}$ is the $m$-vector ${\bf u}^T = \frac{1}{\sqrt{m}}[1, 1,
\ldots, 1]$ and noting that ${\bf u}^T {\bf u} = 1$, we observe
that the only non-zero eigenvalue of the matrix ${\bm \Psi}_{m}$
is equal to $m$. It follows that the only non-zero eigenvalue of
the matrix ${\bf M} {\bf M}^H$ is equal to $ m n \mu^2$, and thus
$\omega_1 = m n  \mu^2$ and $\omega_i = 0$ for $i = 2, \ldots, m$.
Substituting these into (\ref{eq:rc_nonCentrlWishartLatentRoots2})
and using the definition of the Rician factor\index{Rician factor}
$\kappa$ from (\ref{eq:rc_rice factor1}) we have
\begin{eqnarray}
f_{{\bf \lambda}_1, \ldots, {\bf \lambda}_m}(\lambda_1, \ldots,
\lambda_m) &=& (1+ \kappa)^{m n} e^{-m n \kappa} {}_0\tilde{F}_1
(n; (1+ \kappa) {\bm \Omega}, (1+ \kappa) {\bf \Lambda}) \times
\nonumber \\
&& \frac{\pi^{m(m-1)}}{\tilde{\Gamma}_{m}(n)
\tilde{\Gamma}_{m}(m)} e^{- (1+ \kappa) \sum_{i=1}^m \lambda_i}
\left(\prod_{i=1}^m \lambda_i\right)^{n-m} \prod_{i<j}^m \left(
\lambda_i - \lambda_j \right)^2 .
\label{eq:rc_nonCentrlWishartLatentRoots3}
\end{eqnarray}
Note that, when $\kappa = 0$,
(\ref{eq:rc_nonCentrlWishartLatentRoots3}) reduces to the
distribution of the Rayleigh fading latent roots, given in
\cite{Telatar1}, since in this case ${\bm \Omega} = {\bf 0}_{m
\times m}$ and ${}_0\tilde{F}_1 (n; {\bf 0}_{m \times m} , {\bf
\Lambda}) = 1$.

The general capacity expression for the Rician fading channel for
an arbitrary number of transmit/receiver antennas is then given
from (\ref{eq:rc_capacity_Rician4}), by,
\begin{eqnarray}
C_{N_T,N_R} = \int_{\lambda_1, \ldots, \lambda_m} \sum_{i=1}^m
\log \left( 1 + \frac{P}{N_T} \lambda_i \right) f_{{\bf
\lambda}_1, \ldots, {\bf \lambda}_m}(\lambda_1, \ldots, \lambda_m)
d \lambda_1 \ldots d \lambda_m  \ ,
\label{eq:rc_capacity_Rician7}
\end{eqnarray}
where $f_{{\bf \lambda}_1, \ldots, {\bf \lambda}_m}(\lambda_1,
\ldots, \lambda_m)$ is given in
(\ref{eq:rc_nonCentrlWishartLatentRoots3}) above.

\section{A New Signalling Scheme for Multiple Transmit Antenna
Systems in Rician Fading} \label{sec:rc_newSigs}

In Section \ref{sec:rc_RiceCap2_t1} we observed that there is a
large gap between the capacity upper bound for the Rician multiple
transmit antenna system under the assumption of known $\kappa$ at
the transmitter and the capacity of the system without this
assumption. In this section we propose a better signalling scheme
for multiple antenna systems that achieves higher capacity by
explicitly making use of the knowledge of the Rician factor at the
transmitter.

Recall from Section \ref{sec:rc_RiceCap1} that the optimal signal
choice for such a multiple antenna system in Rician fading,
subjected to an average transmit power constraint $P$, is
zero-mean complex Gaussian. Thus, the only thing we do not know is
the covariance matrix structure of the optimal input signal
distribution. Based on the discussion at the beginning of Section
\ref{sec:rc_RiceCap1}, we propose the following choice for the
covariance matrix of the zero-mean complex Gaussian input signal
${\bf x}$:
\begin{eqnarray}
{\bf Q}^{\kappa} &=& \frac{P}{N_T (1+\kappa)} \left( {\bf I}_{N_T}
+ \kappa {\bm \Psi}_{N_T} \right) \label{eq:rc_newQ2}
\end{eqnarray}
where, as before, ${\bm \Psi}_{N_T}$ is the $N_T \times N_T$
matrix of all ones. Note that, when $\kappa =0$,
(\ref{eq:rc_newQ2}) becomes the optimal covariance for the
Rayleigh channel. On the other hand, as $\kappa \longrightarrow
\infty$, ${\bf Q}^{\kappa} \longrightarrow {\bf Q}^{\infty}$ which
is the optimal covariance for the AWGN MIMO channel. Thus ${\bf
Q}^{\kappa}$ reduces to the optimal covariance matrices at these
two extremes.

With this choice for the covariance matrix ${\bf Q}$, the capacity
of the multiple antenna system becomes
\begin{eqnarray}
C_{N_T,N_R}^{\kappa} &=& \emph{E}_{{\bf H}} \left\{ \log \det
\left( \frac{P}{N_T (1+\kappa)}  {\bf H} \left( {\bf I}_{N_T} +
\kappa {\bm \Psi}_{N_T} \right) {\bf H}^H +  {\bf I}_{N_R} \right)
\right\} . \nonumber
\end{eqnarray}
Note that we may write ${\bm \Psi}_{N_T} = {\bf e} {\bf e}^T$
where ${\bf e}$ denotes the $N_T-$vector of all ones. Since the
matrix ${\bf H}$ reduces to an $N_T$ length row vector when
$N_R=1$, the capacity of the multiple antenna system in this case
can be written as
\begin{eqnarray}
C_{N_T,1}^{\kappa} &=& \emph{E}_{Z,W} \left\{ \log \left(
\frac{P}{N_T (1+\kappa)}  \left(  W + \kappa |Z|^2 \right) + 1
\right) \right\} , \label{eq:rc_capacity_Rician_kappa2}
\end{eqnarray}
where as usual $W =  {\bf H} {\bf H}^H = \sum_{n_T=1}^{N_T} |
\left( {\bf H} \right)_{1,n_T}|^2$ and we have defined $Z = {\bf
e}^T {\bf H}^H  = \sum_{n_T=1}^{N_T} \left( {\bf H}
\right)_{1,n_T}$. Since the $ \left( {\bf H} \right)_{1,n_T}$'s
are independent complex ${\cal N}_c \left(
\frac{\mu}{\sqrt{2}}(1+j), 2 \sigma^2 \right)$ random variables,
it follows that $Z \sim {\cal N}_c \left( N_T
\frac{\mu}{\sqrt{2}}(1+j), 2 N_T \sigma^2 \right)$. Hence, using
our earlier notation, it can be easily shown that $|Z|^2 \sim
{\cal W}_1 \left( 1, \frac{N_T}{1+\kappa}, N_T \kappa \right)$ and
$W \sim {\cal W}_1 \left( N_T, \frac{1}{1+\kappa}, N_T \kappa
\right)$.

It is clear that these two random variables $Z$ and $W$ are not
independent and so we do not know their joint distribution, which
is required for evaluating (\ref{eq:rc_capacity_Rician_kappa2}).
Thus we resort to capacity bounds and derive both upper and lower
bounds on the capacity of the multiple antenna system with the
choice (\ref{eq:rc_newQ2}). In particular, we obtain a tight lower
bound on the capacity which shows that the proposed choice of the
covariance matrix is far superior to the scaled identity
covariance matrix for any non-zero $\kappa$ (and, of course, is
the same as that capacity when $\kappa=0$).

\subsection{Upper Bound for $C_{N_T,1}^{\kappa}$}

Applying Jensen's inequality to
(\ref{eq:rc_capacity_Rician_kappa2}) and noting that $\emph{E}
\left\{ |Z|^2 \right\} = \frac{N_T}{1+\kappa} \left( 1 + N_T
\kappa \right)$ and $\emph{E} \left\{ W \right\} = N_T $, we have
the following upper bound on the capacity of a multiple transmit
and single receiver antenna system in Rician fading with the
proposed covariance matrix:
\begin{eqnarray}
C_{N_T,1}^{\kappa} &\leq&   \log \left( 1 + \frac{P}{1+\kappa}
\left( 1+  \frac{\kappa}{1+\kappa} +  \frac{N_T
\kappa^2}{1+\kappa} \right) \right) .
\label{eq:rc_capacity_Rician_kappaUB}
\end{eqnarray}

It can be shown that in the special cases of $\kappa=0$ and
$\kappa \longrightarrow \infty$, the upper bound
(\ref{eq:rc_capacity_Rician_kappaUB}) reduces respectively to,
\begin{eqnarray}
C_{N_T,1}^{\kappa=0} &\leq&   \log \left( 1 + P \right) ,
\label{eq:rc_capacity_Rician_kappaUB1}
\end{eqnarray}
and
\begin{eqnarray}
C_{N_T,1}^{\kappa=\infty} &\leq&   \log \left( 1 + N_T P \right) .
\label{eq:rc_capacity_Rician_kappaUB2}
\end{eqnarray}
From (\ref{eq:rc_capacity_rician_asymp3}) and the results in
\cite{Telatar1} we know that the right-hand side of
(\ref{eq:rc_capacity_Rician_kappaUB1}) is in fact the exact
capacity of the system in this case as $N_T \longrightarrow
\infty$. Hence, in the case of $\kappa=0$ the upper bound
(\ref{eq:rc_capacity_Rician_kappaUB}) is achieved as $N_T
\longrightarrow \infty$. Also, from the remarks in Section
\ref{sec:RicianCapLarge_t} following
(\ref{eq:rc_capacity_rician_asymp4}), we see that the right-hand
side of (\ref{eq:rc_capacity_Rician_kappaUB2}) indeed is the exact
capacity of the system in this case for any value of $N_T$. Hence,
when $\kappa \longrightarrow \infty$ the upper bound
(\ref{eq:rc_capacity_Rician_kappaUB}) is achieved for any value of
$N_T$.

\subsection{Lower Bound for $C_{N_T,1}^{\kappa}$}

Since both $W$ and $Z$ are non-negative random variables we may
obtain the following lower bound on the capacity of a multiple
transmit and single receiver antenna system in Rician fading with
the proposed input covariance matrix:
\begin{eqnarray}
C_{N_T,1}^{\kappa} &\geq& (1+\kappa) \exp(-N_T \kappa)
\int_{0}^{\infty} \log \left( \frac{P \kappa}{1+\kappa} z + 1
\right) \exp (-(1+\kappa)z) I_0 \left( 2 \sqrt{N_T \kappa (1+
\kappa) z} \right) dz . \nonumber
\end{eqnarray}

\begin{figure}[hbt!]
\centering
\includegraphics[width=5in]{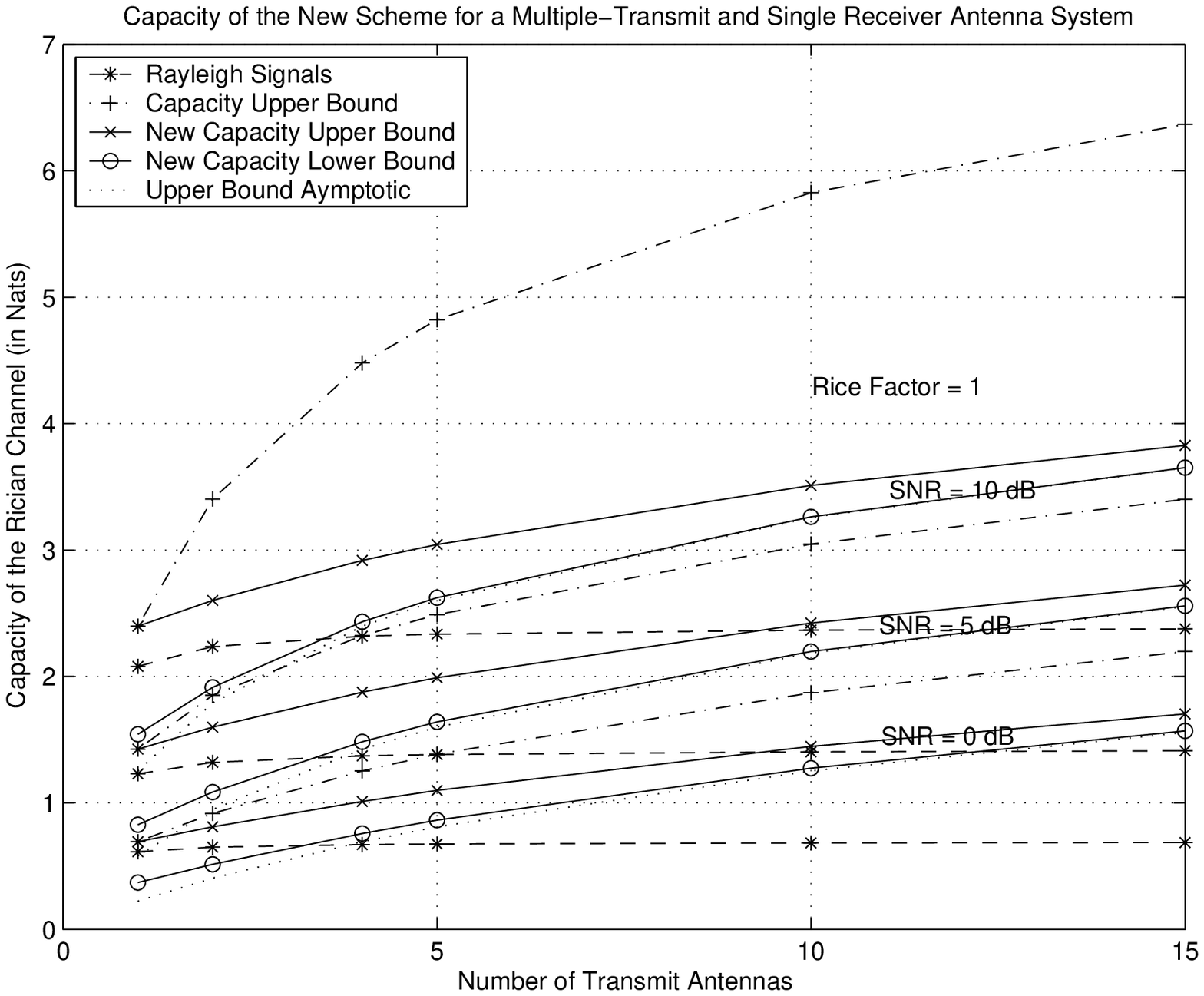}
\caption{Capacity of a Multiple-Transmit Antenna System in Rician
Fading with Proposed New Signalling Scheme. $N_R = 1$ and $\kappa
= 1$. } \label{fig:Cap_n1_r1_newSigs_kappa1}
\end{figure}

\begin{figure}[hbt!]
\centering
\includegraphics[width=5in]{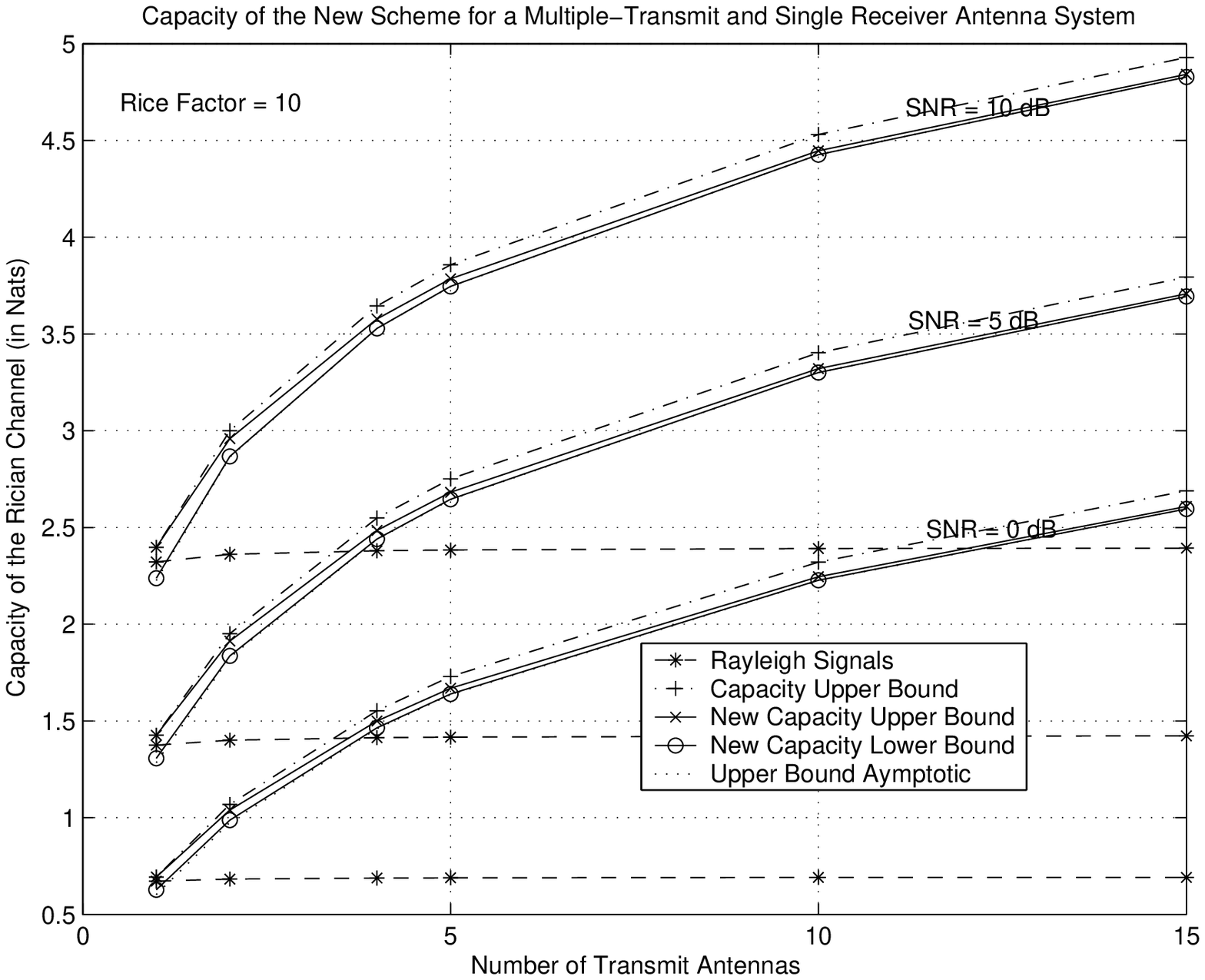}
\caption{Capacity of a Multiple-Transmit Antenna System in Rician
Fading with Proposed New Signalling Scheme. $N_R = 1$ and $\kappa
= 10$. } \label{fig:Cap_n1_r1_newSigs_kappa10}
\end{figure}

In Figs. \ref{fig:Cap_n1_r1_newSigs_kappa1} and
\ref{fig:Cap_n1_r1_newSigs_kappa10} we have shown the derived
bounds for the capacity of a multiple-transmit antenna system
along with the capacity corresponding to the scaled identity
covariance matrix. From the lower bounds shown on these figures it
is clear that the proposed signalling scheme achieves much higher
capacity than that of the scaled identity covariance matrix for
sufficiently large values of $\kappa$ or $N_T$. It is also
observed that the difference between the upper and lower bounds
decreases as $N_T$ increases. From Fig.
\ref{fig:Cap_n1_r1_newSigs_kappa10} we note that when $\kappa$ is
large the upper and lower bounds are almost the same unless the
number of transmit antennas is very small. Thus, for large
$\kappa$, a reasonable approximation to the capacity of the
proposed scheme can be obtained by taking the large-$\kappa$
asymptote of the upper bound
(\ref{eq:rc_capacity_Rician_kappaUB}),
\begin{eqnarray}
C_{N_T,1}^{\kappa} &\approx&   \log \left( 1 + \frac{P N_T
\kappa^2}{(1+\kappa)^2} \right) \ \ \ {\textrm{for}} \ \ \ \kappa
\gg 1 . \label{eq:rc_capacity_Rician_kappaUBasympt}
\end{eqnarray}
In Figs. \ref{fig:Cap_n1_r1_newSigs_kappa1} and
\ref{fig:Cap_n1_r1_newSigs_kappa10} we have also shown this large
$\kappa$ approximation to the upper bound of the capacity. We
observe that indeed (\ref{eq:rc_capacity_Rician_kappaUBasympt}) is
a very good approximation to the capacity even for relatively
small values of $\kappa$. Note also that
(\ref{eq:rc_capacity_Rician_kappaUBasympt}) gives the exact
capacity in the cases of $\kappa=0$ and $\kappa \longrightarrow
\infty$.

Finally, it is worth noting that in these figures we have also
included the capacity upper bound for a Rician channel with
receiver CSI derived in Section \ref{sec:rc_RiceCap1}. From Fig.
\ref{fig:Cap_n1_r1_newSigs_kappa1} we observe that for relatively
small values of $\kappa$ there is still a significant gap between
the general upper bound for the Rician channel in this case given
by (\ref{eq:rc_capacity_RicianGeneral9}) and the upper bound on
the capacity of the proposed new design given by
(\ref{eq:rc_capacity_Rician_kappaUB}). However, as $\kappa$
increases we observe from Fig. \ref{fig:Cap_n1_r1_newSigs_kappa10}
that this difference also becomes smaller, although the proposed
scheme still does not achieve the upper bound
(\ref{eq:rc_capacity_RicianGeneral9}).

\section{Conclusions} \label{sec:rc_conclusions}

In this paper we have investigated the average-power-constrained
capacity of multiple antenna systems under Rician distributed
fading when the receiver has access to channel state information,
but not the transmitter. We have considered two different cases
concerning the knowledge of the fading available at the
transmitter: that in which the transmitter has no knowledge of the
fading at all; and that in which the transmitter has knowledge of
the Rician factor $\kappa$ but not the exact value of CSI. While
obtaining the exact capacity in the former case, we were able to
derive lower and upper bounds for the latter case. The exact
capacity in the former case also quantifies the capacity variation
of a multiple antenna system designed to be optimal for a Rayleigh
fading channel but in fact operating in a Rician fading
environment. For this case, we derived an integral expression for
the capacity of a general system having an arbitrary number of
transmit/receive antennas. In some special cases, we numerically
evaluated this capacity expression. We specifically investigated
the capacity of such systems for large numbers of transmitter
antennas and when only one end of the system (either transmitter
or the receiver) is equipped with a multiple antenna array.

A new signalling scheme has been proposed for the case when the
transmitter knows the value of $\kappa$, though not the exact CSI.
We have analyzed the capacity of this new scheme, in terms of
lower and upper bounds, for a multiple transmit antenna system,
and demonstrated that it offers much higher capacity than that of
the unknown-$\kappa$ capacity-achieving distribution. We also
derived a simple approximation to the capacity of this scheme for
sufficiently large values of $\kappa$.

Our results indicate that Rician fading can improve the capacity
of a multiple antenna system, especially if the transmitter knows
the value of $\kappa$. Moreover, the proposed signalling scheme
provides a mechanism for capturing this improvement.

\appendix
In this appendix we present a few mathematical concepts that have
been used throughout this paper. Most of these are related to
various types of special functions needed in our analysis.

The complex multivariate gamma function, $\tilde{\Gamma}_m(a)$, is
defined as \cite{James1}
\begin{eqnarray}
\tilde{\Gamma}_m(a) &=&  \int_{A^H=A>0} e^{-{\textrm
tr}[A]}|A|^{a-m} (dA) \ = \ \pi^{\frac{1}{2}m(m-1)} \prod_{k=1}^m
\Gamma \left(a-(k-1) \right) . \label{eq:rc_multiGammaComplex2}
\end{eqnarray}
Note that, it follows from (\ref{eq:rc_multiGammaComplex2}) that
$\tilde{\Gamma}_1(a) =   \Gamma (a)$.

The generalized hypergeometric function
\cite{Erdelyi1,James1,Seaborn1} is defined as
\begin{eqnarray}
{}_pF_q (a_1, \ldots, a_p;b_1, \ldots, b_q;x) &=&
\sum_{k=0}^{\infty} \frac{(a_1)_k \ldots (a_p)_k}{(b_1)_k \ldots
(b_q)_k} \frac{x^k}{k!} \label{eq:rc_ac_hypergeometricFunc1}
\end{eqnarray}
where $p$ and $q$ are integers, and the hypergeometric coefficient
$(a)_k$ is defined as the product
\begin{eqnarray}
(a)_k &=& a (a+1) \ldots (a+k-1) ,
\label{eq:rc_HyperGeoMetricCoeff1}
\end{eqnarray}
with $(a)_0 = 1$.

The complex hypergeometric function ${}_p\tilde{F}_q (a_1, \ldots,
a_p;b_1, \ldots, b_q;S)$ of an $n \times n$ Hermitian matrix $S$
can be defined as \cite{James1}
\begin{eqnarray}
{}_p\tilde{F}_q (a_1, \ldots, a_p;b_1, \ldots, b_q;S) &=&
\sum_{k=0}^{\infty} \sum_{\kappa} \frac{(a_1)_{\kappa} \ldots
(a_p)_{\kappa}}{(b_1)_{\kappa} \ldots (b_q)_{\kappa}}
\frac{\tilde{C}_{\kappa}(S)}{k!} \nonumber
\end{eqnarray}
where $p$ and $q$ are integers, $\tilde{C}_{\kappa}(S)$ is the
{\it zonal polynomial}
(\cite{James3,James4,Muirhead2,Subrahmaniam1}) of the $n \times n$
Hermitian matrix $S$ corresponding to the partition $\kappa =
(k_1, k_2, \ldots, k_n), \ k_1 \geq k_2 \geq \ldots \geq k_n \geq
0$, of the integer $k$ into not more than $n$ parts and
$[a]_{\kappa}$ is the complex multivariate hypergeometric
coefficient defined as
\begin{eqnarray}
[a]_{\kappa} &=& \prod_{i=1}^{n} { \left( a-(i-1) \right) }_{k_i}
. \nonumber
\end{eqnarray}
Note that $\tilde{C}_{\kappa}(S)$  is a homogeneous symmetric
polynomial of degree $k$ in the latent roots of the matrix $S$.

The hypergeometric functions with two argument matrices, $S$ and
$T$ (both $n \times n$), can then be defined via the relation
\cite{Constantine1,James1,Muirhead2},
\begin{eqnarray}
{}_p\tilde{F}_q(a_1, \ldots, a_p;b_1, \ldots, b_q;S,T) &=&
\int_{{\cal U}(n)} {}_p\tilde{F}_q(a_1, \ldots, a_p;b_1, \ldots,
b_q;S U T U^H) (dU) \nonumber
\end{eqnarray}
where ${\cal U}(n)$ is the unitary group of all $n \times n$
complex unitary matrices $U$, i.e. $UU^H = I_n$ and $(dU)$ is the
invariant (Haar) measure on ${\cal U}(n)$ normalized to make the
total measure unity.

A special case that we need is ${}_0\tilde{F}_1$, which is the
Bessel function of matrix argument \cite{Bochner1,Herz1,James1},
which can also be represented as
\begin{eqnarray}
{}_0\tilde{F}_1(m;HH^H) &=& \int_{{\cal U}(m)} e^{{\textrm {tr}}(H
U_1 + \overline{H U_1})} (dU) \label{eq:rc_BesselFuncOfMatrix1}
\end{eqnarray}
where $H$ is an $n \times m$ complex matrix with $n \leq m$, $ U =
[U_1, U_2] $ with $U_1$ being  an $n \times m$ complex matrix and
$U_1^H U_1 = I_n$. Note that, in (\ref{eq:rc_BesselFuncOfMatrix1})
$\overline{H}$ and $(dU)$ denote the complex conjugate of the
matrix $H$ and the normalized invariant measure on the unitary
group ${\cal U}(n)$, respectively.

It can be shown that for a scalar $r$,
(\ref{eq:rc_BesselFuncOfMatrix1}) reduces to
\begin{eqnarray}
{}_0\tilde{F}_1(m;r^2) &=& \Gamma(m) r^{-(m-1)} I_{m-1}(2r)
\label{eq:rc_BesselFuncOfMatrixSpecial1}
\end{eqnarray}
where  $I_{\nu}(z)$ is the $\nu$-th order modified Bessel function
of the first kind \cite{McLachlan1}.

\bibliographystyle{IEEEtran} 
\bibliography{XBib}

\begin{thebibliography}{10}
\providecommand{\url}[1]{#1}
\def\UrlFont{\rmfamily}
\providecommand{\newblock}{\relax}
\providecommand{\bibinfo}[2]{#2}
\providecommand\BIBentrySTDinterwordspacing{\spaceskip=0pt\relax}
\providecommand\BIBentryALTinterwordstretchfactor{4}
\providecommand\BIBentryALTinterwordspacing{\spaceskip=\fontdimen2\font plus
\BIBentryALTinterwordstretchfactor\fontdimen3\font minus
  \fontdimen4\font\relax}
\providecommand\BIBforeignlanguage[2]{{%
\expandafter\ifx\csname l@#1\endcsname\relax
\typeout{** WARNING: IEEEtran.bst: No hyphenation pattern has been}%
\typeout{** loaded for the language `#1'. Using the pattern for}%
\typeout{** the default language instead.}%
\else
\language=\csname l@#1\endcsname
\fi
#2}}

\bibitem{Foschini1}
G.~J. Foschini, ``Layered space-time architecture for wireless comunnication in
  a flat fading environment when using multi-element antennas,'' \emph{Bell
  Labs. Tech. Journ.}, vol.~1, no.~2, pp. 41--59, 1996.

\bibitem{FoschiniGans1}
G.~J. Foschini and M.~J. Gans, ``On limits of wireless communications in a
  fading environment when using multiple antennas,'' \emph{Wireless Personal
  Communications}, vol.~6, pp. 311--335, 1998.

\bibitem{Telatar1}
I.~E. Telatar, ``Capacity of multi-antenna {G}aussian channels,'' \emph{Eur.
  Trans. Telecommun.}, vol.~10, pp. 585--595, Nov. 1999.

\bibitem{Stuber1}
G.~L. St\"{u}ber, \emph{Principles of Mobile Communication}.\hskip 1em plus
  0.5em minus 0.4em\relax Norwell, MA: Kluwer Academic Publishers, 1996.

\bibitem{McLachlan1}
N.~W. McLachlan, \emph{Bessel Functions for Engineers}.\hskip 1em plus 0.5em
  minus 0.4em\relax Oxford, UK: Oxford Univ. Press., 1955.

\bibitem{Constantine1}
A.~G. Constantine, ``Some non-central distribution problems in multivariate
  analysis,'' \emph{Ann. Math. Stat.}, vol.~34, pp. 1270--1285, Dec. 1963.

\bibitem{James1}
A.~T. James, ``Distributions of matric variates and latent roots derived from
  normal samples,'' \emph{Ann. Math. Stat.}, vol.~35, pp. 475--501, June 1964.

\bibitem{James2}
------, ``The non-central {W}ishart distribution,'' in \emph{Proc. Royal Soc.
  London. Series A., Math. and Phys. Sci.}, vol. 229, May 1955, pp. 364--366.

\bibitem{gallager1}
R.~G. Gallager, \emph{Information Theory and Reliable Communication}.\hskip 1em
  plus 0.5em minus 0.4em\relax New York: Wiley, 1968.

\bibitem{James3}
A.~T. James, ``The distribution of the latent roots of the covariance matrix,''
  \emph{Ann. Math. Stat.}, vol.~31, pp. 151--158, 1960.

\bibitem{James6}
------, ``The distribution of noncentral means with known covariance,''
  \emph{Ann. Math. Stat.}, vol.~32, pp. 874--882, Sep. 1961.

\bibitem{Erdelyi1}
A.~Erdelyi, Ed., \emph{Higher Transcendental Functions}.\hskip 1em plus 0.5em
  minus 0.4em\relax New York: McGraw-Hill., 1953.

\bibitem{Seaborn1}
J.~B. Seaborn, \emph{Hypergeometric Functions and Their Applications}.\hskip
  1em plus 0.5em minus 0.4em\relax New York: Springer-Verlag, 1991.

\bibitem{James4}
A.~T. James, ``Zonal polynomials of the real positive symmetric matrices,''
  \emph{Ann. of Math.}, vol.~74, pp. 456--469, 1961.

\bibitem{Muirhead2}
R.~J. Muirhead, \emph{Aspects of Multivariate Statistical Theory}.\hskip 1em
  plus 0.5em minus 0.4em\relax New York: Wiley, 1982.

\bibitem{Subrahmaniam1}
K.~Subrahmaniam, ``Recent trends in multivariate distribution theory: {O}n the
  zonal polynomials and other functions of matrix argument. {P}art {I}: zonal
  polynomials,'' Univ. of Manitoba, Tech. Rep.~69, 1974.

\bibitem{Bochner1}
S.~Bochner, ``Bessel functions and modular relations of higher type and
  hyperbolic differential equations,'' \emph{Lunds Univ. Matematiska Seminarium
  Supplementband dedicated to Marcel Riesz}, pp. 12--20, July 1952.

\bibitem{Herz1}
C.~S. Herz, ``Bessel functions of matrix argument,'' \emph{Annals of Math.},
  vol.~61, no.~3, pp. 474--423, May 1955.

\end{thebibliography}

\pagebreak

\pagebreak

\pagebreak

\pagebreak

\pagebreak

\pagebreak

\pagebreak

\pagebreak

\pagebreak

\end{document}